\documentclass[pdftex,twocolumn,epjc3]{svjour3}          

\journalname{Eur. Phys. J. C}
\usepackage{cite}
\usepackage{graphicx}
\usepackage{amssymb} 
\usepackage{amsmath} 
\usepackage{times}
\usepackage{comment}
\usepackage[utf8]{inputenc}

\usepackage{float} 
\usepackage[caption=false]{subfig} 

\usepackage[usenames,dvipsnames]{color}
\usepackage[normalem]{ulem} 
\renewcommand\sout{\bgroup \color[rgb]{0.55,0.00,0.99} \ULdepth=-.5ex \ULset}
\newcommand\soutgr{\bgroup \color{green} \ULdepth=-.5ex \ULset}
\newcommand\soutre{\bgroup \color{red} \ULdepth=-.5ex \ULset}

\newcommand{\sT}{{\scriptscriptstyle T}}
\renewcommand{\d}{\mathrm{d}}
\def\slash#1{\setbox0=\hbox{$#1$}               
        \dimen0=\wd0                            
        \setbox1=\hbox{/} \dimen1=\wd1          
        \ifdim\dimen0>\dimen1                   
        \rlap{\hbox to \dimen0{\hfil/\hfil}}    
        #1                                      
        \else
        \rlap{\hbox to \dimen1{\hfil$#1$\hfil}} 
        /                                       
        \fi}                                    %

\usepackage[normalem]{ulem} 
\renewcommand\sout{\bgroup \color[rgb]{0.55,0.00,0.99} \ULdepth=-.5ex \ULset}

\newcommand{\bme}[1]{\mbox{\boldmath ${\scriptstyle #1}$}}

\usepackage{graphicx}
\usepackage{amssymb} 
\usepackage{amsmath} 
\usepackage{comment}
\usepackage[utf8]{inputenc}
\usepackage{lineno}

 \usepackage{ifpdf}
 \ifpdf
 \usepackage[pdftex]{hyperref}
 \else
 \usepackage[hypertex]{hyperref}
 \fi

 \hypersetup{
   pdftitle={},%
   pdfauthor={},%
   pdfsubject={},%
   pdfkeywords={},%
   pdfstartview={},%
   bookmarksopen=true, breaklinks=true, debug=true, %
   colorlinks=true, linkcolor=red, citecolor=blue, urlcolor=blue
 }

\usepackage{float} 
\usepackage[caption=false]{subfig} 

\newcount\savefnused
\newcount\savefndone

\newcommand{\savefootnote}[2][\empty]
{\ifx\empty#1\footnotemark\else\footnotemark[#1]\fi
 \global\advance\savefnused by 1
 \expandafter\xdef\csname savefnmark\the\savefnused\endcsname{\thefootnote}%
 \expandafter\xdef\csname savefntext\the\savefnused\endcsname{#2}%
}
\newcommand{\flushfootnote}{\loop\ifnum\savefndone<\savefnused
  \global\advance\savefndone by 1
  \footnotetext[\csname savefnmark\the\savefndone\endcsname]%
    {\csname savefntext\the\savefndone\endcsname}%
  \global\expandafter\let\csname savefnmark\the\savefndone\endcsname\relax
  \global\expandafter\let\csname savefntext\the\savefndone\endcsname\relax
\repeat}

\usepackage{multirow,bigdelim}
\usepackage{tabularx,ragged2e}%

\newcolumntype{Y}{>{\centering\arraybackslash}X}

\newcommand{\nn}{\nonumber}

\def\GeV{{\rm GeV}}
\def\GeV2{{\rm GeV}^2}

\newcommand{\Q}{{\cal Q}}

\newcommand{\CS}{{\rm CS}}

\newcommand{\mc}[1]{\mathcal{#1}}
\newcommand{\kT}{\bm k_{\sT}}
\newcommand{\koneT}{\bm k_{1\sT}}
\newcommand{\ktwoT}{\bm k_{2\sT}}
\newcommand{\qT}{\boldsymbol{P}_{\Q\Q\sT}}

\renewcommand{\d}{\mathrm{d}}

\newcommand{\ce}[1]{Eq.~(\ref{#1})}
\newcommand{\cf}[1]{{Fig.~\ref{#1}}}

\newcommand{\bm}[1]{\mbox{\boldmath $#1$}}

\newcommand{\Cff}{$\mathcal{C}\Big[f_1^{\, g}f_1^{\, g}\Big]$}
\newcommand{\Cwtwohh}{$\mathcal{C}\Big[w_2\,h_1^{\perp\, g}h_1^{\perp\, g}\Big]$}
\newcommand{\Cwthreefh}{$\mathcal{C}\Big[w_3\,f_1^{\, g}h_1^{\perp\, g}\Big]$}
\newcommand{\Cwfourhh}{$\mathcal{C}\Big[w_4\,h_1^{\perp\, g}h_1^{\perp\, g}\Big]$}

 \usepackage{ifpdf}
 \ifpdf
 \usepackage[pdftex]{hyperref}
 \else
 \usepackage[hypertex]{hyperref}
 \fi

 \hypersetup{
   pdftitle={},%
   pdfauthor={},%
   pdfsubject={},%
   pdfkeywords={},%
   pdfstartview={},%
   bookmarksopen=true, breaklinks=true, debug=true, %
   colorlinks=true, linkcolor=red, citecolor=blue, urlcolor=blue
 }

\begin{document} 

\title{{Studies of gluon TMDs and their evolution using quarkonium-pair production at the LHC}}

\author{
Florent Scarpa\thanksref{addr1,addr2} 
\and 
Dani\"el Boer\thanksref{addr1}
\and 
Miguel G. Echevarria\thanksref{addr3,addr4}
\and
Jean-Philippe Lansberg\thanksref{addr2} 
\and
Cristian Pisano\thanksref{addr5}
\and
Marc Schlegel\thanksref{addr6}
}
\institute{Van Swinderen Institute for Particle Physics and Gravity,
University of Groningen, Nijenborgh 4, 9747 AG Groningen, The Netherlands\label{addr1}
\and
IPNO, CNRS-IN2P3, Univ. Paris-Sud, Universit\'e Paris-Saclay, 
91406 Orsay Cedex, France\label{addr2}
\and
Istituto Nazionale di Fisica Nucleare, Sezione di Pavia, via Bassi 6, 27100 Pavia, Italy\label{addr3}
\and
Dpto. de F\'isica y Matem\'aticas, Universidad de Alcal\'a,
Ctra. Madrid-Barcelona Km. 33, 28805 Alcal\'a de Henares (Madrid), Spain\label{addr4}
\and
Dipartimento di Fisica, Universit\`a di Cagliari, and INFN, Sezione di Cagliari,
    Cittadella Universitaria, I-09042 Monserrato (CA), Italy\label{addr5}
\and
Department of Physics, New Mexico State University, Las Cruces, NM 88003, USA\label{addr6}
}


\date{Version of \today}

\maketitle

\begin{abstract}
$J/\psi$- or $\Upsilon$-pair production at the LHC are promising processes to study the gluon transverse momentum distributions (TMDs) which remain very poorly known. In this article, we improve on previous results by including the TMD evolution in the computation of the observables such as the pair-transverse-momentum spectrum and asymmetries arising from the linear polarization of gluons inside unpolarized protons. We show that the azimuthal asymmetries generated by the gluon polarization are reduced compared to the tree level case but are still of measurable size (in the 5\%-10\% range). Such asymmetries should be measurable in the available data sets of $J/\psi$ pairs and in the future data sets of the high-luminosity LHC for $\Upsilon$ pairs.

\end{abstract}

\section{Introduction}

The three-dimensional structure of the composite hadrons has widely been analyzed through the study of Transverse-Momentum Dependent parton distribution functions (TMDs) in the framework of TMD factorization. 
The various TMDs can be accessed in hadronic processes with a small transverse momentum (TM), denoted by $q_T$, of the detected final state~\cite{Ralston:1979ys,Sivers:1989cc,Tangerman:1994eh}. 
TMDs need to be extracted from experimental data for such processes as they are intrinsically nonperturbative objects and therefore cannot be computed using perturbative QCD. 
So far, the majority of data allowing for the extraction of TMDs have been acquired from SIDIS and Drell-Yan measurements, two experimentally accessible processes and for which TMD factorization was proved to hold \cite{Collins:2011zzd,GarciaEchevarria:2011rb,Echevarria:2012js}. 
However, since such processes are primarily induced by quarks/antiquarks, they mostly provide information about the {\it quark} TMDs. 
Currently our knowledge of {\it gluon} TMDs is still very limited, due to the lack of data on processes that could potentially be used for extractions. 
More specifically, gluons inside unpolarized protons can be described at leading twist using two TMDs \cite{Mulders:2000sh}. 
The first one describes unpolarized gluons, while the second one describes linearly polarized gluons. 
The latter correlates the spin of the gluons with their TM, and thus requires non-zero gluon TM. 
The presence of polarized gluons inside the unpolarized proton has effects on the cross-sections, such as modifications of the TM-spectrum and azimuthal asymmetries.

Several processes have been proposed to extract gluon TMDs, see e.g.~\cite{Boer:2011kf,Sun:2011iw,Qiu:2011ai,Godbole:2012bx,Boer:2012bt,Godbole:2013bca,Godbole:2014tha,Zhang:2014vmh,Boer:2014lka,Dunnen:2014eta,Boer:2014tka,Zhang:2015yba,Mukherjee:2015smo,Echevarria:2015uaa,Mukherjee:2016qxa,Mukherjee:2016cjw,Boer:2016bfj,Lansberg:2017tlc,Godbole:2017syo,DAlesio:2017rzj,Rajesh:2018qks,Bacchetta:2018ivt,Lansberg:2017dzg,Kishore:2018ugo,Sun:2012vc,Ma:2012hh,Ma:2014oha,Ma:2015vpt}.
Associated quarkonium production (see~\cite{Lansberg:2019adr} for a recent review) has in particular a great potential to probe the gluon TMDs at the LHC, e.g.\ quarkonium plus photon ($\mathcal{Q} + \gamma$) or quarkonium-pair production.
They mainly originate from gluon fusion, and can be produced via a color-singlet transition, avoiding then possible TMD-factorization-breaking effects \cite{Collins:2007nk,Collins:2007jp,Rogers:2010dm}.

Some quarkonium states, like the $J/\psi$ meson, are easily detected and a large number of events can be recorded. 
Processes with two particles in the final state offer some interesting advantages compared to those with a single detected particle. 
Since the TM of the final state needs to be small for the cross-section to be sensitive to TMD effects, one-particle final states are bound to stay close to the beam axis and therefore difficult to detect, as the background level is high and triggering is complicated. 
However, two particles that are nearly back-to-back can each have large individual transverse momenta that add up to a small one. 
Indeed, in general a pair of particles can have a large invariant mass and a small TM.
Whereas the hard scale in a one-particle final state is only its mass, and is thus constant, the invariant mass of a two-particle final state can be tuned with their individual momenta. This allows one to study the scale evolution of the TMDs. 
Finally, a two-particle final state allows one to define the azimuthal angle between these two particles, hence to look for various azimuthal asymmetries. 
These are in fact associated to specific convolutions of gluon TMDs.

It was thus recently proposed~\cite{Lansberg:2017dzg} to probe the gluon TMDs using quarkonium-pair production at the LHC, and more specifically $J/\psi + J/\psi$ production. 
Such a process has already been measured by LHCb, CMS and ATLAS at the LHC, as well as by D0 at the Tevatron~\cite{Abazov:2014qba,Khachatryan:2014iia,Aaij:2011yc,Aaij:2016bqq,Aaboud:2016fzt}. 
The size of some azimuthal asymmetries associated with the linearly polarized gluon distribution are nearly maximum in this process. 
In~\cite{Lansberg:2017dzg}, the unpolarized-gluon distribution was modelled by a simple Gaussian as a function of the gluon TM. 
In order to see the maximal effect of the linearly-polarized gluons on the yields, their distribution was taken to saturate its positivity bound~\cite{Mulders:2000sh}. 
The size of the resulting maximum asymmetries was found to be very large, especially at large pair invariant mass, $M_{\Q\Q}$. 
Yet, more realistic estimates of the asymmetries require the inclusion of higher-order corrections in $\alpha_s$ through TMD QCD evolution~\cite{Collins:2011ca,Echevarria:2012pw,Echevarria:2014rua,Echevarria:2015uaa}.

Very recently, a TMD-factorization proof has been established for pseudoscalar $\eta_{c,b}$ hadro-production at low TM~\cite{Echevarria:2019ynx}. 
To date, this is the only one for quarkonium hadroproduction. 
It was pointed out that new hadronic matrix elements are involved for quarkonium production at low TM, in addition to the TMDs. 
These encode the soft physics of the process.
It is not known how much these new hadronic matrix elements impact the phenomenology. 
In this context, we build on the previous work~\cite{Lansberg:2017dzg} by adding TMD evolution effects to the gluon TMDs.
Such evolution effects are expected to play a significant role (see e.g.\ \cite{Echevarria:2015uaa}) and should in any case be specifically analyzed.
We will proceed like in previous studies for $H^0$ production~\cite{Sun:2011iw,Boer:2014tka,Echevarria:2015uaa}.

In this article, we first discuss the characteristics of quarkonium-pair production at the LHC within the TMD framework, as well as the associated cross-section and observables sensitive to the gluon TMDs. 
We then detail the evolution formalism used in our computations and the resulting expressions for the TMD convolutions. 
Finally, we present our results for the $\qT$-spectrum and the azimuthal asymmetries for $J/\psi$-pair production at the LHC as well as azimuthal asymmetries for $\Upsilon$-pair production.

\section{$\Q$-pair production within TMD factorization}

\subsection{TMD factorization description of the process}

\begin{figure}[hbt!]
\centering
\includegraphics[width=\columnwidth]{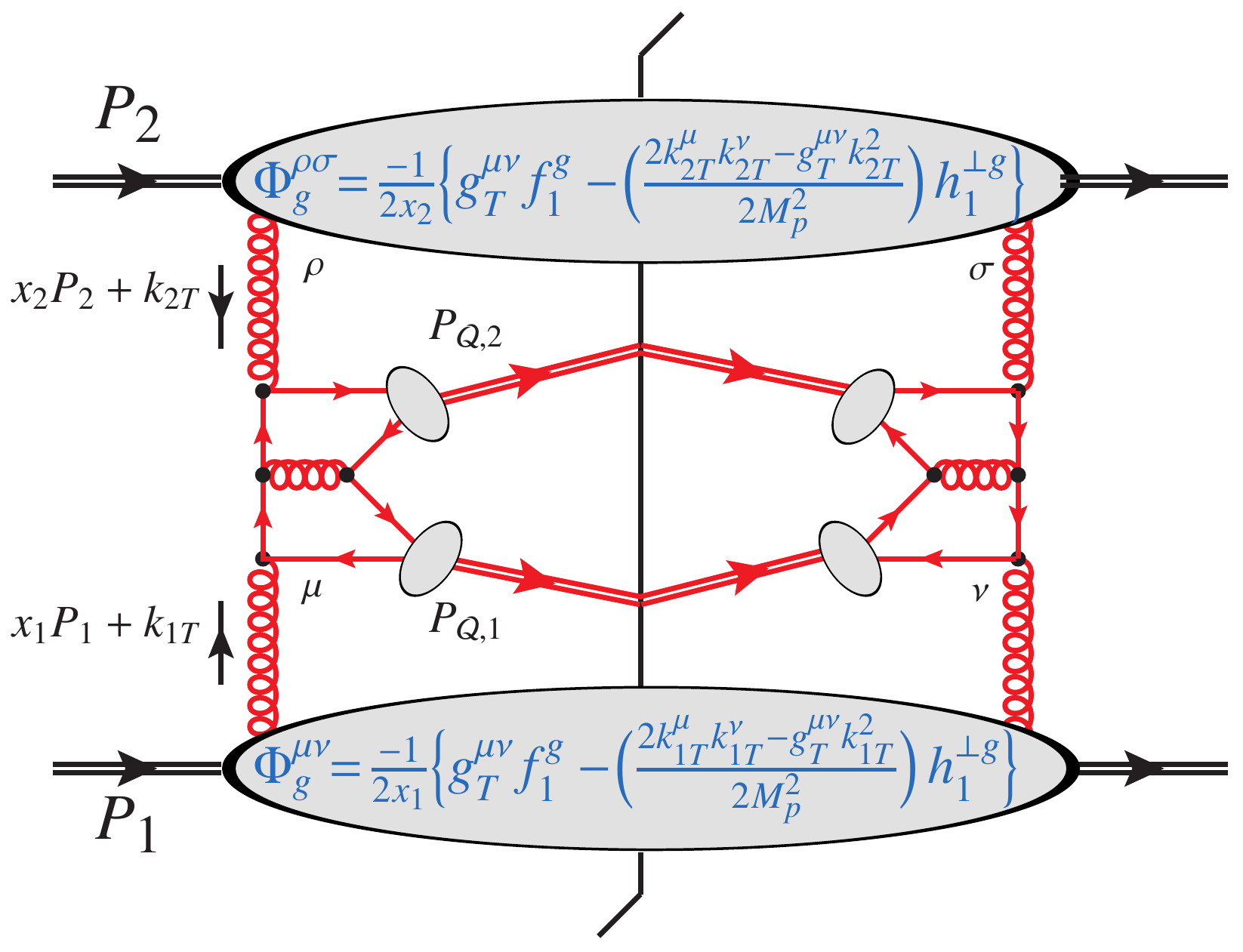}
\vspace*{-0.5cm}
\caption{Representative Feynman graph for $p(P_1) {+} p(P_2) \to {\Q} (P_{\Q,1}) {+}  {\Q} (P_{\Q,2}) {+}X$ via 
gluon fusion at LO in  TMD factorisation.
}\vspace*{-0.25cm}
\label{fig:Feynman-graph}
\end{figure}

TMD factorization extends collinear factorization by taking into account the intrinsic TM of the partons, usually denoted by $\kT$. 
As in collinear factorization, the hard-scattering amplitude, which can be perturbatively computed, is multiplied by parton correlators that can be parametrized in terms of parton distribution functions, but in this case $\kT$ dependent.  
The parametrization of parton correlators is an extension from that used in collinear factorization, not only because of the $\kT$ dependence of the distribution functions, but also because there are more distributions.
The gluon correlator inside an unpolarized proton with momentum $P$ and mass $M_p$, denoted by $\Phi_g^{\mu\nu}(x,\kT)$ \cite{Mulders:2000sh,Meissner:2007rx,Boer:2016xqr}, can be parametrized in terms of two independent TMDs. 
The first one is the distribution of unpolarized gluons $f_1^{\,g}(x,\kT^2)$, the second one is the distribution of linearly polarized gluons $h_1^{\perp\,g}(x,\kT^2)$. 
Here the gluon 4-momentum is written using a Sudakov decomposition: $k = xP + k_{\sT} + k^-n$ (where $n$ is any light-like vector ($n^2=0$) such that $n\cdot P \neq 0$), where $\bm k_\sT^2 = -k_\sT^2$ and the transverse metric is $g^{\mu\nu}_{\sT} = g^{\mu\nu} - (P^{\mu} n^{\nu}+ P^\nu n^\mu)/P{\cdot} n$. 
For TMD factorization to hold, the hard scale of the process should be much larger than the pair TM, $q_T$.

The process we are interested in is the fusion of two gluons coming from two colliding unpolarized protons, leading to the production of a pair of vector $S$-wave quarkonia: $g(k_1) {+} g(k_2) \to  \Q(P_{\Q,1}) {+}  \Q(P_{\Q,2})\,$. 
The cross-section for this reaction involves the contraction of two gluon correlators \cite{Lansberg:2017dzg}, $\Phi_g^{\mu\nu}(x_1,\koneT)$ and $\Phi_g^{\rho\sigma}(x_2,\ktwoT)$, with the squared amplitude $\mathcal{M}^{\mu\rho}(\mathcal{M}^{\nu\sigma})^*$ of the partonic scattering, integrated over the gluon momenta.
The expression of the tree-level partonic amplitude $\mathcal{M}$ is available in \cite{Qiao:2009kg}, although the earliest computations date back to 1983 \cite{Kartvelishvili:1984ur,Humpert:1983yj}. 
The hadronization process, i.e.\ the transition from a heavy-quark pair to a quarkonium bound state, is described in our study using the color singlet model (CSM)~\cite{Chang:1979nn,Baier:1981uk,Baier:1983va} or in this case equivalently non-relativistic QCD (NRQCD) \cite{Bodwin:1994jh} at LO in the velocity $v$ of the heavy quarks in the bound-state rest frame. 
Fig.\ \ref{fig:Feynman-graph} represents the complete reaction with a typical Feynman diagram depicting the partonic subprocess.
 
\subsection{Other contributions to quarkonium-pair production}

The leading contribution to the hadronization of a $Q\overline{Q}$ pair into a bound state in NRQCD is the Color-Singlet (CS) transition, for which the perturbatively-produced heavy-quark pair has the same quantum numbers as the quarkonium and directly binds without any extra soft interaction. 
Corrections to this leading contribution involving higher Color-Octet (CO) Fock states are suppressed by powers of $v$, which is meant to be much smaller than unity for heavy quarkonia.

The CS over CO dominance normally follows from this power suppression in $v$ encoded in the so-called NRQCD long distance matrix elements (LDMEs). 
More precisely one expects a relative suppression on the order
of $v^4$~\cite{Bodwin:1994jh,Cho:1995ce,Cho:1995vh} (see~\cite{Lansberg:2019adr,Andronic:2015wma,Brambilla:2010cs,Lansberg:2006dh} for reviews) per quarkonium. 
For di-$J/\psi$ production with $v^2_c\simeq 0.25$ and for which both the CO and the CS yields are produced at $\alpha_s^4$, the CO/CS yield ratio, which thus scales as $v_c^8$, likely lies below the percent level. 
Explicit computations~\cite{Ko:2010xy,Li:2013csa,Lansberg:2014swa,He:2015qya,Lansberg:2019fgm} indeed show corrections from the CO states below the percent level except in some corners of the phase space (e.g.\ large rapidity separation $\Delta y$) where some CO contributions can be kinematically enhanced, but these can safely be avoided with appropriate kinematical cuts. More details can be found in~\cite{Lansberg:2019fgm}.

It is important for the applicability of TMD factorization that the CS contributions dominate. 
Soft gluon interactions between the hadrons and a colored initial or final state of the hard scattering can be encapsulated within the definition of the TMD through the use of Wilson lines. 
However, if both initial and final states are subject to soft gluon interactions, the resulting color entanglement may break TMD factorization~\cite{Collins:2007nk,Collins:2007jp,Rogers:2010dm}. 
The dominance of the CS contributions should therefore be ensured.

It is also important to take into account $\alpha_s$ corrections. 
In the TMD region, $\qT\ll M_{\Q\Q}$, these introduce a renormalization scale ($\mu$) dependence in the TMD correlators and a  rapidity scale $\zeta$ dependence~\cite{Collins:2011zzd,GarciaEchevarria:2011rb,Echevarria:2012js}. 
At larger $\qT$ one has to match onto the collinear factorization expression (see e.g.~\cite{Echevarria:2018qyi}), which is calculated by taking real-gluon emissions into account~\cite{Lansberg:2013qka,Lansberg:2014swa,Sun:2014gca,Likhoded:2016zmk}. 
At finite $\qT$, such single real-gluon emissions occur at $\alpha_s^5$ and the quarkonium pair effectively recoils against this hard gluon, increasing the pair TM. 
In this paper we will restrict to  $\qT < M_{\Q\Q}/2$ in order to stay away from the matching region. 

Thus far, we have focused our discussion on the single parton scattering (SPS) case. 
However, since we look at a two-particle final state, we should also consider the case where the quarkonia are created in two separate hard scatterings, i.e.~double parton scattering (DPS). 
At LHC energies, the gluon densities are typically high and the likelihood for two hard gluon fusions to take place during the same proton-proton scattering cannot be neglected. 

In the case of di-$J/\psi$, it has been already anticipated in 2011~\cite{Kom:2011bd} that DPS contributions may be dominant at large rapidity difference $\Delta y$ (thus large invariant masses with same individual TM). 
This was corroborated~\cite{Lansberg:2014swa} by the CMS data~\cite{Khachatryan:2014iia} with an excess above the SPS predictions at large $\Delta y$. 
ATLAS further~\cite{Aaboud:2016fzt} confirmed the DPS relevance in di-$J/\psi$ production with a dedicated DPS study.
One expects\footnote{Theory DPS studies advance~\cite{Diehl:2018kgr,Buffing:2017mqm,Diehl:2017kgu,Diehl:2017wew},  but not yet as to provide {\it quantitative} inputs to predict DPS cross {sections} as done for SPS. As such, one usually assumes the DPS {contributions} to be independent.  This justifies factorizing the DPS cross section into  individual ones with an (inverse) proportionality factor, referred to as an effective cross-section $\sigma_{{\rm eff}}$. Under this assumption, $\sigma_{{\rm eff}}$ should be process independent, encoding the magnitude of the parton interaction. $\sigma_{{\rm eff}}$ needs to be experimentally extracted as it is a nonperturbative quantity. This is the standard procedure at LHC energies~\cite{Akesson:1986iv,Alitti:1991rd,Abe:1993rv,Abe:1997xk,Abazov:2009gc,Aad:2013bjm,Chatrchyan:2013xxa}. 
Ideally, a single precise extraction of $\sigma_{{\rm eff}}$ should suffice to provide predictions for any DPS cross section under this factorized Ansatz. 
Yet, the current extraction seems to differ~\cite{Lansberg:2017chq} with values ranging from 25~mb down to a few mb, which forces us to restrict to qualitative considerations.}~\cite{Lansberg:2014swa,Aaboud:2016fzt} that DPS contributions lie below 10\% for $\Delta y \sim 0$ in the CMS and ATLAS samples (characterized by a $P_{\Q T}$ cut away from the threshold $M_{\psi\psi}\simeq 2M_\psi$) and that they only matter at large $\Delta y$.
However, in the LHCb acceptance (where $M_{\psi\psi}\simeq 2M_\psi$), DPS contributions cannot a priori be neglected, but can be subtracted~\cite{Aaij:2016bqq} if one assumes that, for the DPS sample, the kinematics of both $J/\psi$'s is uncorrelated. This yield a precise (yet, unnormalized) prediction of the kinematical distributions. 
 
Another source of quarkonium pairs is the feed-down from excited states. 
For $J/\psi$-pair production, the main feed-down sources are the $\chi_c$ and $\psi'$. 
The feed-down from $\chi_c$ is expected to be small, as $gg \rightarrow J/\psi + \chi_c$ and $gg \rightarrow \psi' + \chi_c$ are suppressed~\cite{Lansberg:2014swa} at LO by $C$-parity and the vanishing of the $\chi_c$ wave function at the origin, and $gg \rightarrow \chi_c + \chi_c$  is suppressed by the squared $\chi_c \rightarrow J/\psi$ branching ratio.
The production of a $\psi'$ with a $J/\psi$ likely contributes~\cite{Lansberg:2019adr} 50\% of the $J/\psi$-pair samples, owing to the large branching ratio for $\psi'\to J/\psi$ (${\cal O}(60)\%$) and symmetry factors. 
Yet, $\psi'+J/\psi$ pairs are produced exactly like $J/\psi+J/\psi$ pairs and thus generate the same TMD observables~\footnote{Up to the small kinematical shift due to the decay which we neglect in what follows.}. 

In the case of $\Upsilon$-pair production, the main feed-down is from $\Upsilon(2,\!3S)$. 
According to \cite{Berezhnoy:2012tu}, less than 30\% of the produced pairs would originate from feed-down at $\sqrt{s}$ = 8~TeV.
As in the $J/\psi$ case, $C$-parity suppresses the $\Upsilon+\chi_b$ reaction at leading order, and $\chi_b+\chi_b$ is suppressed by the squared branching ratio. 
Regarding the CO contributions, the relative velocity of the quarks inside the $\Upsilon$ is smaller than for the $J/\psi$, meaning the NRQCD expansion used to describe the hadronization has a better convergence. 
Therefore it is highly unlikely that the CO channels overcome the CS ones in the reachable phase space. 
The fraction of DPS events is also expected to be less than 5\% at low $\qT$ and central rapidity \cite{Lansberg:2019adr}, making it an overall cleaner process.

\subsection{The TMD differential cross-section}

The general structure of the TMD-based differential cross-section describing quarkonium-pair production from gluon fusion reads~\cite{Lansberg:2017dzg}: 
\begin{align}\label{eq:crosssection} 
&\frac{\d\sigma}{\d M_{\Q\Q} \d Y_{\Q\Q} \d^2 \qT \d \Omega} =
  \frac{\sqrt{M_{\Q\Q}^2 - 4 M_\mc{Q}^2}}{(2\pi)^2 8 s\, M_{\Q\Q}^2}
   \\
  \Bigg\{	
  &F_1(M_{\Q\Q},\theta_{\CS})\ \mc{C} \Big[f_1^gf_1^g\Big](x_{1,2},\qT)  \nn \\
  + &F_2(M_{\Q\Q},\theta_{\CS})\ \mc{C} \Big[w_2h_1^{\perp g}h_1^{\perp g}\Big](x_{1,2},\qT)
  \nn \\
  + &
  \Bigg(F_3(M_{\Q\Q},\theta_{\CS}) \ \mc{C} \Big[w_3 f_1^g h_1^{\perp g}\Big](x_{1,2},\qT)+  \nn \\
   & F'_3(M_{\Q\Q},\theta_{\CS})\ \mc{C} \Big[w'_3 h_1^{\perp g} f_1^g \Big](x_{1,2},\qT)\Bigg)  \cos2\phi_{\CS} \ 
  \nn \\
  + & 
   F_4(M_{\Q\Q},\theta_{\CS})\ \mc{C}\! \left[w_4 h_1^{\perp g}h_1^{\perp g}\right](x_{1,2},\qT) \cos 4\phi_{\CS} 
  \!\Bigg \} \nn
  \,,
\end{align}
with $\d\Omega=\d\!\cos\theta_{\CS}\d\phi_{\CS}$, $\{\theta_{\CS},\phi_{\CS}\}$ being the Collins-Soper (CS) angles \cite{Collins:1977iv}, and
$Y_{\Q\Q}$  is the rapidity of the pair. $x_{1,2} = M_{\Q\Q}\,e^{\pm Y_{\Q\Q}}/\sqrt{s}$, with $s = (P_1 + P_2)^2$.
Here $\qT$ ($\equiv q_T$) and $Y_{\Q\Q}$ are defined in the hadron c.m.s. The quarkonia move  along (in the opposite direction) $\vec e=(\sin\theta_{\CS}\cos \phi_{\CS},
\sin\theta_{\CS}\sin \phi_{\CS}, \cos\theta_{\CS})$ in the CS frame. The kinematical pre-factor is 
specific to the mass of the quarkonia and the considered differential cross-sections, while
the hard-scattering coefficients $F_i$ only depend on $\theta_{\CS}$ and the invariant mass of the system, here $M_{\Q\Q}$. Their expression for quarkonium-pair production can be found at tree level in~\cite{Lansberg:2017dzg}.  
When $P_{\Q T} \gg M_\Q$, small values of $\cos\theta_{\CS}$ correspond to small values of $\Delta y$ in the hadron c.m.s. 

 The TMD convolutions appearing in \ce{eq:crosssection} are defined as follows:
\begin{align}\label{eq:Cwfg}
&\mathcal{C}[w\, f\, g](x_{1,2},\qT) \equiv \nn\\
&\int\!\! \d^{2}\koneT\!\! \int\!\! \d^{2}\ktwoT\,
  \delta^{2}(\koneT+\ktwoT-{\qT})\times \nn\\ &w(\koneT,\ktwoT)\, 
  f(x_1,\koneT^{2})\, 
  g(x_2,\ktwoT^{2}) 
  \,,
\end{align}
where $w(\koneT,\ktwoT)$ denotes a TMD weight. 
The weights in Eq.\ \eqref{eq:crosssection} are common to all gluon-fusion processes originating from unpolarized proton collisions. 
They can be found in \cite{Lansberg:2017tlc}. 
Our aim in the present study is to study the impact of QCD evolution effects in the above TMD convolutions. 
Having at our disposal the computation of the hard-scattering coefficients, the measurements of differential yields in principle allow one to extract these TMD convolutions evolved up to the natural scale of the process, on the order of $M_{\Q\Q}$ here.

In practice, one looks at specific observables sensitive to these convolutions.
First we note that when the cross-section is integrated over the azimuthal angle $\phi_{\CS}$ , the terms with a $\cos(2,\!4\phi_{\CS})$-dependence drop out from Eq.\ \eqref{eq:crosssection} such that
\begin{align}
\frac{1}{2\pi}&\int \!\!d\phi_{\CS} \frac{d\sigma}{d M_{\Q\Q} d Y_{\Q\Q} d^2 \qT d \Omega} =
\nn\\
&F_1\, \mc{C} \Big[f_1^{\,g}f_1^{\,g}\Big]+F_2\, \mc{C} \Big[w_2h_1^{\perp\, g}h_1^{\perp\, g}\Big]
\,,
\end{align}
giving direct access to $\mc{C} \Big[f_1^{\,g}f_1^{\,g}\Big]$ and $\mc{C} \Big[w_2h_1^{\perp\, g}h_1^{\perp\, g}\Big]$.

Furthermore, one can define, at fixed $\{Y,\qT,\theta_{\CS},M_{\Q\Q}\}$, $\cos(n\phi_{\CS})$-weighted differential cross-sections, integrated over $\phi_{\CS}$ and normalized by their azimuthally-independent component:
\begin{align}
&\langle  \cos(n\phi_{\CS}) \rangle = \nn\\
&\frac{\displaystyle \int \!\!d\phi_{\CS} \cos(n\phi_{\CS})\,  \frac{\displaystyle\d\sigma}{d M_{\Q\Q} d Y_{\Q\Q} d^2 \qT d \Omega}}{\displaystyle\!\!\int \!\!d\phi_{\CS} \frac{d\sigma}{d M_{\Q\Q} d Y_{\Q\Q} d^2 \qT d \Omega}}\, .
\end{align}
Such a variable, computed for $n$ = 2 or 4 in our case, corresponds to (half of) the relative size of the $\cos(2,\!4\phi_{\CS})$-modulations present in the TMD cross-section in comparison to its  $\phi_{\CS}$-independent component:
\begin{eqnarray}\label{asym_exp}
\langle\cos 2\phi_{\CS}\rangle & = & \frac{1}{2}\;\frac{F_3 \mc{C} \Big[w_3 f_1^{\,g} h_1^{\perp\, g}\Big] + F'_3 \mc{C} \Big[w'_3 h_1^{\perp\, g} f_1^{\,g} \Big]}{F_1\, \mc{C} \Big[f_1^{\,g}f_1^{\,g}\Big]+F_2\, \mc{C} \Big[w_2h_1^{\perp\, g}h_1^{\perp\, g}\Big]} \, ,\nn\\
\langle\cos 4\phi_{\CS}\rangle & = & \frac{1}{2}\;\frac{F_4 \mc{C}\! \left[w_4 h_1^{\perp\, g}h_1^{\perp\, g}\right]}{F_1\, \mc{C} \Big[f_1^{\,g}f_1^{\,g}\Big]+F_2\, \mc{C} \Big[w_2h_1^{\perp\, g}h_1^{\perp\, g}\Big]}\, .
\end{eqnarray}

When $\langle \cos n\phi_{\CS} \rangle$ is computed within a range of $M_{\Q\Q}$, $Y_{\Q\Q}$, $\qT$ or $\cos(\theta_{\CS})$, we define it as the ratio of corresponding integrals. Of course, the range in $\qT$ should be such that one remains in the TMD region, i.e.\ $\qT\ll M_{\Q\Q}$.  

For positive Gaussian $h_1^{\perp\, g}$ the $\langle \cos(2\phi_{CS}) \rangle$ asymmetry will be positive (note that in~\cite{Lansberg:2017dzg} the $\langle \cos(2\phi_{CS}) \rangle$ plots miss an overall minus sign).

\section{TMD evolution formalism}

TMD evolution has been considered in an increasing number of TMD observables. 
It is usually implemented by Fourier transforming to $b_T$-space, with $b_T$ being the conjugate variable to $\qT$.
When evolution effects are considered, the TMDs acquire a dependence on two scales: a renormalization scale $\mu$ and a rapidity scale $\zeta$ (whose evolution is governed by the Collins-Soper equation).
Below we present in a simple way the results needed to perform the TMD evolution.
For more details, we refer to e.g.\ \cite{Collins:2011ca,Echevarria:2012pw,Echevarria:2014rua,Echevarria:2015uaa}.

When TMD evolution is incorporated to the gluon TMDs in the tree-level result in Eq.~\eqref{eq:crosssection}, the convolutions take the form
\begin{align}\label{eq:Cwfg2}
\mathcal{C}&[w\, f\, g](x_{1,2},\qT;\mu) \equiv \nn \\ 
&\int\!\! \d^{2}\koneT\!\! \int\!\! \d^{2}\ktwoT\,
  \delta^{2}(\koneT+\ktwoT-{\qT})\times  \nn \\
  & w(\koneT,\ktwoT)\,f(x_1,\koneT^{2};\zeta_1,\mu)\, 
  g(x_2,\ktwoT^{2};\zeta_2,\mu) 
  \,,
\end{align}
where the two rapidity scales should fulfill the constraint $\zeta_1 \zeta_2=M_{\Q\Q}^4$.
While the renormalization scale $\mu$ in the hard-scattering coefficients $F_i$ should be set here to $\mu\sim M_{\Q\Q}$ in order to avoid large logarithms, the TMDs should be evaluated at their natural scale $\mu\sim\sqrt{\zeta}\sim \mu_b=b_0/b_T$ (with $b_0=2e^{-\gamma_E}$), in order to minimize both logarithms of $\mu b_T$ and $\zeta b_T^2$, and then evolved up to $\mu\sim\sqrt{\zeta}\sim M_{\Q\Q}$.
The solution of the evolution equations results in the introduction of the following Sudakov factor $S_A$:
\begin{eqnarray}\label{Cf1f1_mub}
\tilde{f}_1^{\,g}(x_1,b_T^2;\zeta,\mu) & = & 
e^{-\frac{1}{2}S_A(b_T;\zeta,\mu)}\tilde{f}_1^{\,g}(x,b_T^2;\mu_b^2,\mu_b)
\,,
\nn\\
\tilde{h}_1^{\perp\, g}(x_1,b_T^2;\zeta,\mu) & = & 
e^{-\frac{1}{2}S_A(b_T;\zeta,\mu)}\tilde{h}_1^{\perp\, g}(x,b_T^2;\mu_b^2,\mu_b)
\end{eqnarray}
where the Fourier-transformed TMDs are
\begin{eqnarray}
\tilde{f}_1^{\,g}(x,\bm b_T^2;\zeta,\mu) & = & 
\int\!d^2\bm k_T\, e^{-i\bme b_T \cdot \,\bme k_T}f_1^{\,g}(x,\bm k_T^2;\zeta,\mu)
\,,
\nn\\
\tilde{h}_1^{\perp\, g}(x,\bm b_T^2;\zeta,\mu) & = & 
\int\!d^2\bm k_T\, \frac{(\bm b_T\cdot\,\bm k_T)^2-\frac{1}{2}\bm b_T^2 \bm k_T^2}{\bm b_T^2 M_p^2} \nn\\ 
&&\times e^{-i\bme b_T\cdot\,\bme k_T }h_1^{\perp\, g}(x,\bm k_T^2;\zeta,\mu) 
\,,
\end{eqnarray}
and the perturbative Sudakov factor (applicable for sufficiently small $b_T$) is given by
\begin{align}
S_A(b_T;\zeta,\mu)&=
2D(\mu_b^2)\,\ln\frac{\zeta}{\mu_b^2}
+\\
&2\int_{\mu_b}^{\mu}\!\!\frac{d\bar\mu}{\bar\mu}
\Bigg[ \Gamma(\alpha_s(\bar\mu^2))\ln\frac{\zeta}{\bar\mu^2} 
+ \gamma(\alpha_s(\bar\mu^2))\Bigg]\nn
\,.
\end{align}
We consider here the resummation at next-to-leading-logarithmic accuracy, for which the Collins-Soper kernel $D$ and the non-cusp anomalous dimension $\gamma$ need to be taken at leading-order, while the cusp anomalous dimension $\Gamma$ at next-to-leading-order. 
The perturbative Sudakov factor then takes the form
\begin{align}
&S_A(b_T;\zeta,\mu)= 
2\frac{C_A}{\pi} \int_{\mu_b}^{\mu}\!\!\frac{d\bar\mu}{\bar\mu}\ln\left(\frac{\zeta}{\bar\mu^2}\right) \\ &
\times \Bigg[ \alpha_s(\bar\mu^2)   
+
\bigg(\big(\frac{67}{9}-\frac{\pi^2}{3}\big)
-\frac{20T_fn_f}{9}\bigg)\frac{\alpha_s^2(\bar\mu^2)}{4\pi} \Bigg]
\nn\\
&\quad
+ 2\frac{C_A}{\pi}\int_{\mu_b}^{\mu}\!\!\frac{d\bar\mu}{\bar\mu}
\alpha_s(\bar\mu^2)\left[ -\frac{11-2n_f/C_A}{6} \right]
\,,
\end{align}
with $C_A=3$, $T_f=1/2$ and $n_f$ the number of flavors (we will use $n_f=4$ for di-$J/\psi$ and $n_f=5$ for di-$\Upsilon$ production).
The running of $\alpha_s$ is implemented at one loop.
We note that the Sudakov factor $S_A$ is spin independent, and thus the same for all (un)polarized TMDs \cite{Echevarria:2014rua,Echevarria:2015uaa}.

The perturbative component of the TMDs for small $b_T$ can be computed at a given order in $\alpha_s$. 
At leading order, $\tilde{f}_1^{\,g}$ is given by the integrated PDF:
\begin{equation}
\label{f1pert}
\tilde{f}_1^{\,g}(x,b_T^{2};\zeta,\mu)  =  
f_{g/P}(x;\mu)
+\mathcal{O}(\alpha_s)
+\mathcal{O}(b_T\Lambda_{\rm QCD})
\,.
\end{equation}
As said above, $h_1^{\perp\, g}$ describes the correlation between the gluon polarization and its TM ($k_T$) inside the unpolarized proton. 
It requires a helicity flip and therefore an additional gluon exchange. 
Consequently, its perturbative expansion starts at $\mathcal{O}(\alpha_s)$~\cite{Sun:2011iw}:
\begin{align}
\label{h1pert}
&\tilde{h}_1^{\,\perp g}(x,b_T^{2};\zeta,\mu)  =  -\!\left(
\frac{\alpha_s(\mu)C_A}{\pi}\!\!\int_x^1\!\!\frac{d\hat{x}}{\hat{x}}\!\!\left(\frac{\hat{x}}{x}-1\right)\!\! f_{g/P}(\hat{x};\mu)\right.
\nn\\
&
+\frac{\alpha_s(\mu)C_F}{\pi}\sum_{i=q,\bar q}\int_x^1\!\!\left.\frac{d\hat{x}}{\hat{x}}\left(\frac{\hat{x}}{x}-1\right)f_{i/P}(\hat{x};\mu)\right) \nn\\
&+\mathcal{O}(\alpha_s^2)
+\mathcal{O}(b_T\Lambda_{\rm QCD})
\,,
\end{align}
The above equations in principle allow one to derive a perturbative expression of these TMDs. However, they are strictly applicable only in a restricted $b_T$ range, whereas we need an expression for them from small to large $b_T$ in order to perform the corresponding Fourier transform.

For large $b_T$, one indeed leaves the domain of perturbation theory. On the contrary, when $b_T$ gets too small, $\mu_b$ becomes larger than $M_{\Q\Q}$ and the evolution should stop. The above perturbative expression for the Sudakov factor
should thus not be used as it is.

One of the common solutions to continue to use the above expressions consists in replacing $b_T$ by a function of $b_T$ which freezes in both these limits such that one is not sensitive to the physics there. 
For our numerical studies we use the following $b_T$ prescription~\cite{Collins:2016hqq}:
\begin{align}
\label{eq:bprescription}
b_T^*\big(b_c(b_T)\big) &= 
\frac{b_c(b_T)}{\sqrt{1+\Big(\frac{b_c(b_T)}{b_{T_{\max}}}\Big)^2}}
\end{align}
where 
\begin{align}
b_c(b_T	) = \sqrt{b_T^2+\Big(\frac{b_0}{M_{\Q\Q}}\Big)^2}
\,,
\end{align}
such that $\mu_b=\frac{b_0}{b_T^*(b_c)}$ always lies between $b_0/b_{T_{\max}}$ (reached when $b_T\rightarrow \infty$) and $M_{\Q\Q}$ (reached when $b_T\rightarrow 0$). 
This prescription is of course not unique, as it entails e.g.\ some particular assumptions on the transition from the hard to the soft regime.
The ambiguity in the choice of this prescription can however be absorbed in the nonperturbative modelling of the TMDs, anyhow needed in the large $b_T$ region, which we discuss next.

Schematically each TMD convolution can be written in $b_T$-space as
\begin{equation}
\mathcal{C}[w\, f\, g] = \int_0^\infty\!\! \frac{db_T}{2\pi}\, b_T^n J_m(b_T q_T)\,\tilde{W}(b_T,Q)\,,
\end{equation}
for some integers $n$ and $m$. Here $\tilde W$ is a simple product of Fourier-transformed TMDs. 
The nonperturbative Sudakov factor $S_{{\rm NP}}$ is now defined through 
$\tilde{W}(b_T,Q)=\tilde{W}(b_T^*,Q)\, e^{-S_{{\rm NP}}(b_T,Q)}$, where by construction $\tilde{W}(b_T^*,Q)$ is perturbatively calculable for all $b_T$ values.
The value of $b_{T\max}$ in Eq.~\eqref{eq:bprescription} (roughly) sets the separation between the perturbative and nonperturbative domains. Its optimal value depends on many factors, such as the functional form chosen for $b_T^*$ and the parametrization of the nonperturbative Sudakov factor $S_{{\rm NP}}$. 
For our numerical studies we take $b_{T\max}=1.5$ GeV$^{-1}$, inspired by previous fits from Drell-Yan and $W, Z$ production~\cite{Landry:2002ix,Konychev:2005iy,Aybat:2011zv,Collins:2014jpa,DAlesio:2014mrz}. 

The functional form of $S_{{\rm NP}}$ has been subject of debate, but is usually chosen to be proportional to $b_T^2$ for all $b_T$. By definition $e^{-S_{{\rm NP}}(b_T,Q)}$ has to be equal to 1 for $b_T=0$ and for large $b_T$ it has to vanish, at the very least to ensure convergence of the results. It is usually assumed to be a monotonically decreasing function of $b_T$ and its change from 1 to 0 is assumed to happen within the confinement distance. Lacking experimental constraints, here we will assume a simple Gaussian form (of varying widths). In order to assess the importance of the nonperturbative Sudakov factor for the size of the asymmetries and to perform a first error estimate, we consider several functions. 
For this purpose, we take a simple formula for the nonperturbative Sudakov factor that encapsulates the expected $M_{\Q\Q}$-dependence \cite{Collins:1981va} and the assumed $b_T$-Gaussian behavior:
\begin{equation}\label{Snptune}
S_{{\rm NP}}\big(b_c(b_T)\big)=
A \ln\Big(\frac{M_{\Q\Q}}{Q_{\rm NP}}\Big)\, b_c^2(b_T)
\,,
\qquad
Q_{\rm NP}=1~{\rm GeV}
\,.
\end{equation}
{From this nonperturbative Sudakov factors a value $b_{T\lim}$ is defined at which $e^{-S_{{\rm NP}}}$ becomes negligible, to be specific, where it becomes $\sim$ 10$^{-3}$. From this we furthermore define a corresponding characteristic radius $r=\tfrac{1}{2} b_{T\lim}$ (considering $b_{T\lim}$ the diameter, since it is conjugate to $\qT = \bm k_{1T}+\bm k_{2T}$), which delimits the range over which the interactions occur from the center of the proton. To estimate the uncertainty associated with the largely unknown nonperturbative Sudakov factor, we will consider three cases: $b_{T\lim}$ $=$ 2, 4 and 8 GeV$^{-1}$. This spans roughly from $b_{T\max}=1.5$ GeV$^{-1}$ to the charge radius of the proton. The corresponding values of the parameter $A$ and $r$ for $M_{\Q\Q}=12$ GeV are given in Table \ref{table:Avals}.}

\begin{table}[htb]
\centering
\begin{tabular}{c|c|c}
$A$ (GeV$^2$) & $b_{T\lim}$ (GeV$^{-1}$) &  $r$ (fm $\sim 1/(0.2~{\rm GeV}$) \\ \hline
0.64 & 2 &  0.2 \\ 
0.16 & 4 &  0.4 \\ 
0.04 & 8 &  0.8 \\
\end{tabular}
\caption{
Values of the parameter $A$ used in Eq.\ \eqref{Snptune} for $e^{-S_{{\rm NP}}}$, along with the corresponding $b_{T\lim}$ and $r$ at $M_{\Q\Q}=12$ GeV}
\vspace*{-0.25cm}
\label{table:Avals}
\end{table}

\begin{figure}[htb]
\centering
\includegraphics[width=\columnwidth]{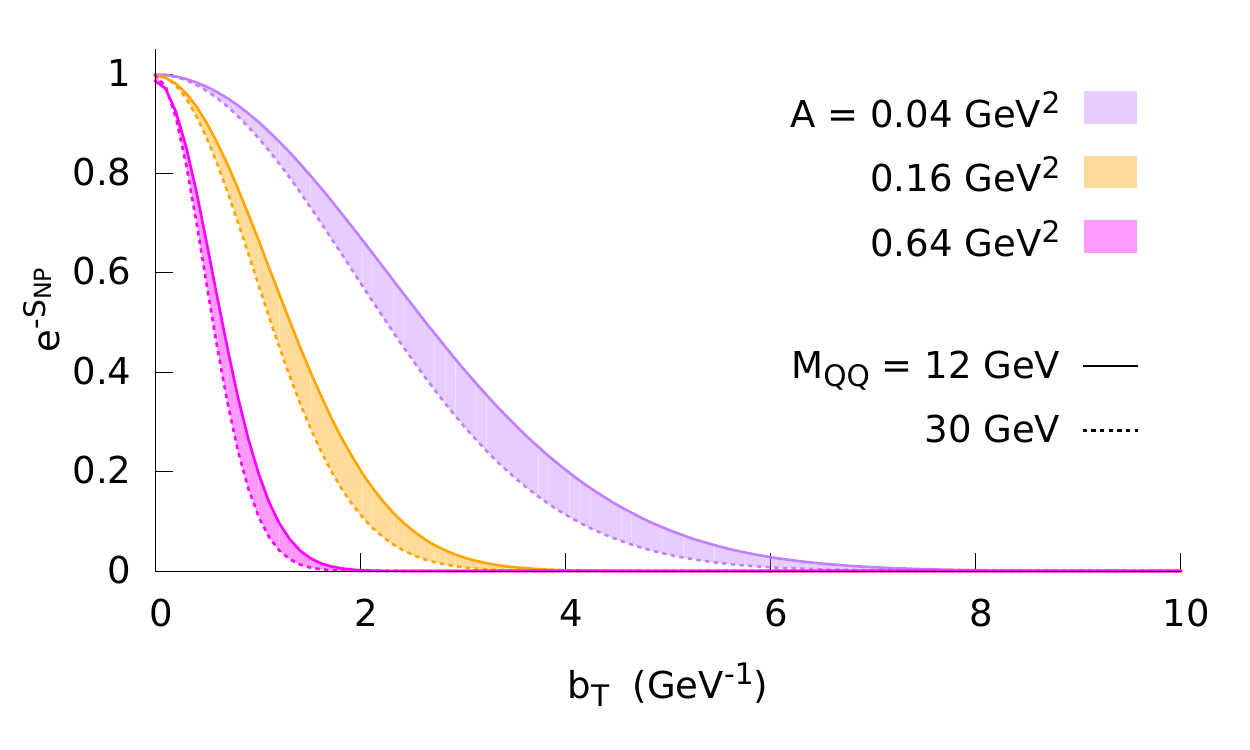}
\vspace*{-0.5cm}
\caption{
$e^{-S_{{\rm NP}}}$ from Eq.\ \eqref{Snptune} vs $b_T$ for $A$ $=$ 0.04 (purple), 0.16 (orange) and 0.64 (magenta) GeV$^2$, for values of $M_{\Q\Q}$ ranging from 12 to 30 GeV. 
The boundaries around the bands depict the exponential at $M_{\Q\Q}$=12 GeV (solid line) and at $M_{\Q\Q}$=30 GeV (dotted line).}
\label{fig:eSnptuneplot}
\end{figure}

The value $M_{\Q\Q}=12$~GeV is considered because the ratio $F_3/F_1$ peaks there (for $J/\psi$ pair production), but we will also consider larger values later on. 
When $M_{\Q\Q}$ increases, the {interaction radius $r$ decreases.} Fig.\ \ref{fig:eSnptuneplot} depicts $e^{-S_{{\rm NP}}}$ as a function of $b_T$ for the three values of $A$ previously mentioned and for $M_{\Q\Q}$ ranging from 12 to 30 GeV.

We point out that the nonperturbative Sudakov factor as fitted by Aybat and Rogers \cite{Aybat:2011zv} to low-energy SIDIS as well as high-energy Drell-Yan and $Z^0$ production data, rescaled by a color factor $C_A/C_F$ to account for the different color representation between quarks and gluons, is very close to the case $b_{T\lim}$ $=$ 2 GeV$^{-1}$. It is also very close to the Fourier transform of the Gaussian model for $f_1^g(x,k_T^2)$ with $\langle k_T^{\, 2} \rangle = 3.3 \pm 0.8$ GeV$^2$ as extracted in \cite{Lansberg:2017dzg} from a LO fit to $J/\psi$-pair-production data from LHCb \cite{Aaij:2016bqq}
from which the DPS contributions was however approximately subtracted. 

We end this section by providing the expressions for the TMD convolutions in $b_T$-space, which we actually use in the numerical predictions in the next section:
\begin{align}
\label{convs_bt}
&\mathcal{C}\Big[f_1^{\, g}f_1^{\, g}\Big] \!=\! 
\int_0^\infty\!\! \frac{db_T}{2\pi}\, b_T J_0(b_T q_T)\, 
e^{-S_A(b_T^*;M_{\Q\Q}^2,M_{\Q\Q})} \,\nn\\& \times e^{- S_{{\rm NP}}(b_c)}
\tilde{f}_1^{\,g}(x_1,b_T^{*\, 2};\mu_b^2,\mu_b)\, 
\tilde{f}_1^{\,g}(x_2,b_T^{*\, 2};\mu_b^2,\mu_b)
\,, 
\nonumber\\
&\mathcal{C}\Big[w_2\,h_1^{\perp\, g}h_1^{\perp\, g}\Big] \!=\! 
\int_0^\infty\!\! \frac{db_T}{2\pi}\, b_T J_0(b_T q_T)\, 
e^{-S_A(b_T^*;M_{\Q\Q}^2,M_{\Q\Q})} \,\nn\\& \times  e^{- S_{{\rm NP}}(b_c)}
\tilde{h}_1^{\perp\, g}(x_1,b_T^{*\, 2};\mu_b^2,\mu_b)\, 
\tilde{h}_1^{\perp\, g}(x_2,b_T^{*\, 2};\mu_b^2,\mu_b) 
\,,
\nonumber\\
&\mathcal{C}\Big[w_3\,f_1^{\, g}h_1^{\perp\, g}\Big] \!=\! 
\int_0^\infty\!\! \frac{db_T}{2\pi}\, b_T J_2(b_T q_T)\, 
e^{-S_A(b_T^*;M_{\Q\Q}^2,M_{\Q\Q})} \,\nn\\& \times  e^{- S_{{\rm NP}}(b_c)}
\tilde{f}_1^{\,g}(x_1,b_T^{*\, 2};\mu_b^2,\mu_b)\, 
\tilde{h}_1^{\perp\, g}(x_2,b_T^{*\, 2};\mu_b^2,\mu_b)
\,, 
\nonumber\\
&\mathcal{C}\Big[w_4\,h_1^{\perp\, g}h_1^{\perp\, g}\Big] \!=\!
\int_0^\infty\!\! \frac{db_T}{2\pi}\, b_T J_4(b_T q_T)\, 
e^{-S_A(b_T^*;M_{\Q\Q}^2,M_{\Q\Q})} \,\nn\\& \times  e^{- S_{{\rm NP}}(b_c)}
\tilde{h}_1^{\perp\, g}(x_1,b_T^{*\, 2};\mu_b^2,\mu_b)\, \tilde{h}_1^{\perp\, g}(x_2,b_T^{*\, 2};\mu_b^2,\mu_b)
\,.
\end{align}

\section{The TM spectrum and the azimuthal asymmetries}

\subsection{$J/\psi$-pair production}

\begin{figure}[htb]
\centering
\subfloat[]{\includegraphics[width=\columnwidth]{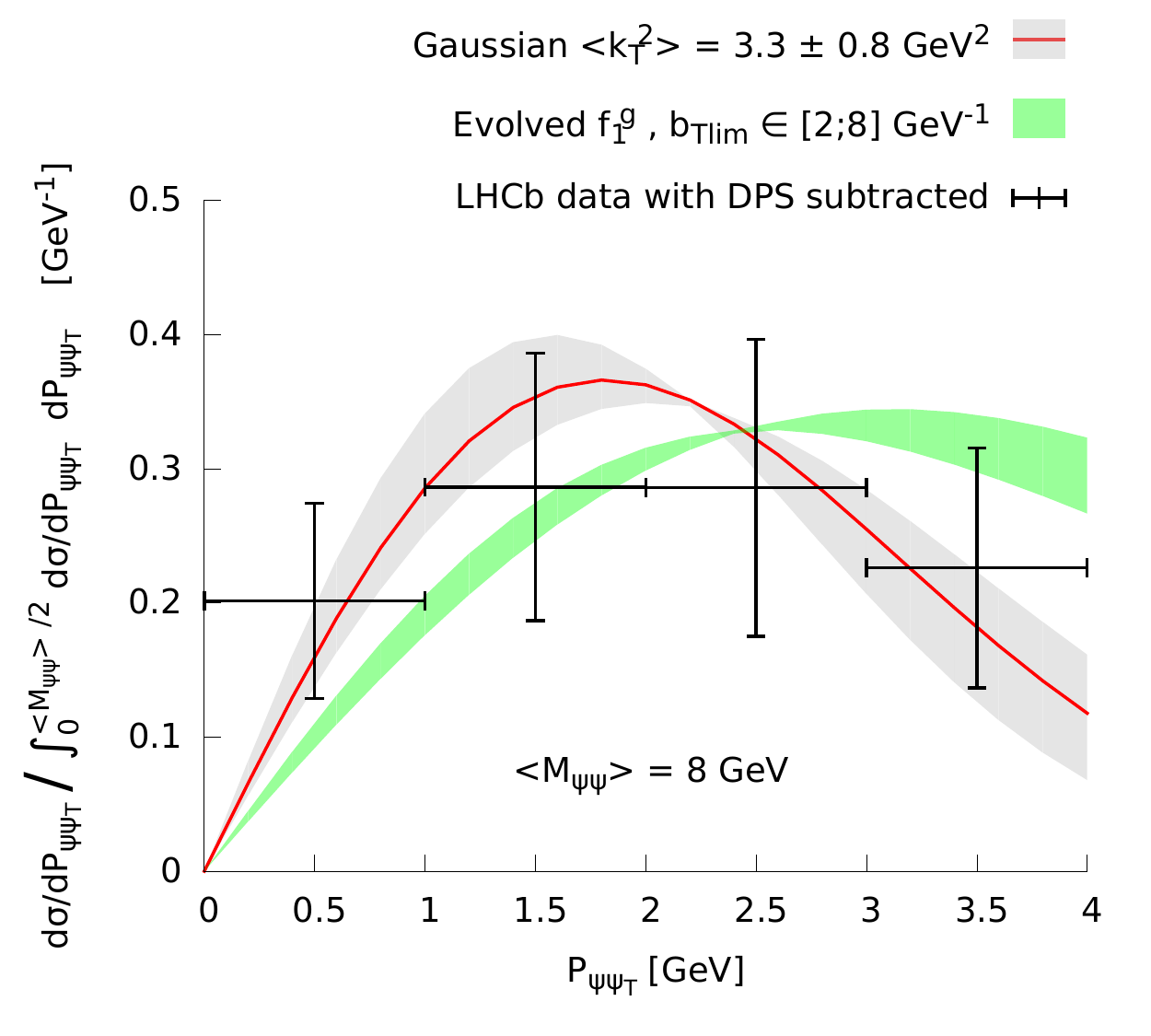}\label{fig:qtCf1f1a}}\\
\subfloat[]{\includegraphics[width=\columnwidth]{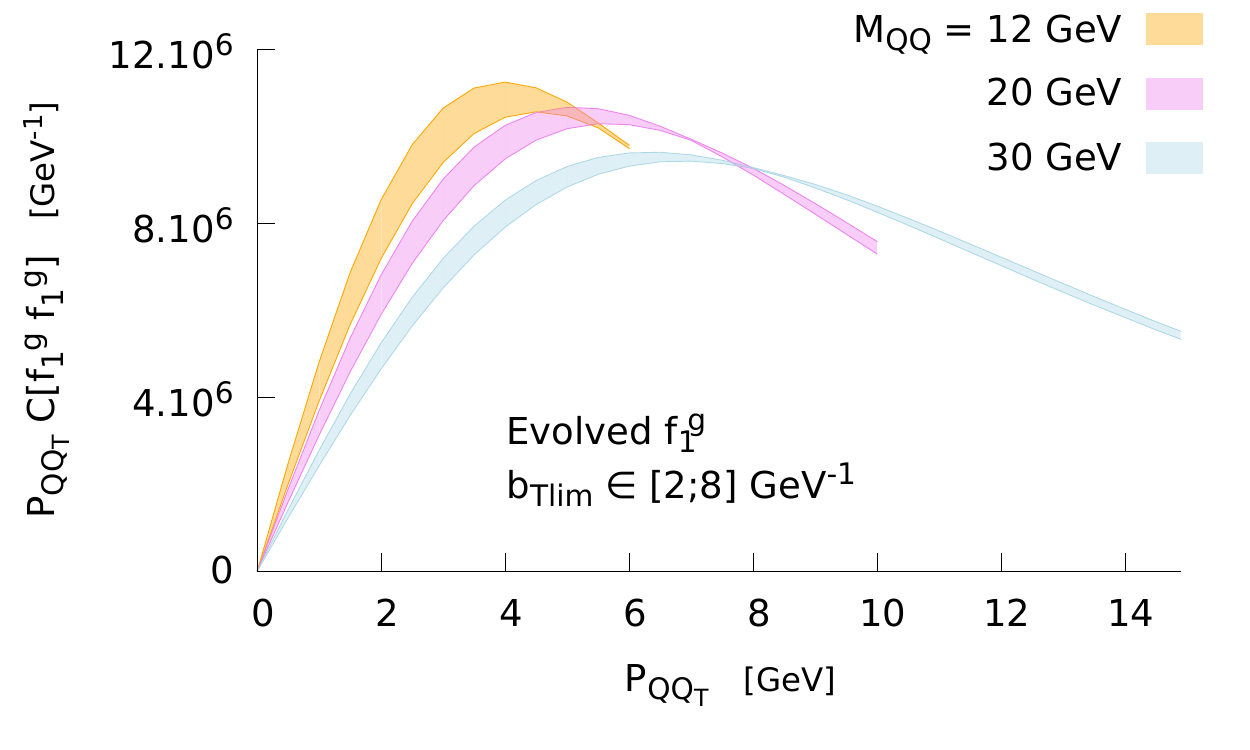}\label{fig:qtCf1f1b}}
\caption{
(a) The \textit{normalised} $\qT$-spectrum for $J/\psi$-pair production at $M_{\psi\psi}$ = 8 GeV using two gluon TMDs. 
The first is a Gaussian Ansatz with $\langle k_T^{\, 2} \rangle = 3.3 \pm 0.8$ GeV$^2$ obtained from the LHCb data~\cite{Lansberg:2017dzg} (the red curve shows the central value and the gray band the associated uncertainty). 
The second is the result of our present study with TMD evolution.
The green band results from the uncertainty on the $b_T$-width of the nonperturbative Sudakov factor $S_{{\rm NP}}$. The estimated DPS contribution has been subtracted from the LHCb data (black crosses) which were also normalized over the interval.
(b) The $\qT$-spectrum using our evolved gluon TMDs at $M_{\Q\Q}$ = 12, 20 and 30 GeV for the same uncertainty on the $b_T$-width.}
\label{fig:qtCf1f1}
\end{figure}

As said, after integration over the azimuthal angle $\phi_{CS}$, one gets to a good approximation $d\sigma/dq_T\propto q_T\, \mathcal{C}[f_1^{\, g}f_1^{\, g}]$. 
In Fig.\ \ref{fig:qtCf1f1} (a) we compare $q_T\, \mathcal{C}[f_1^{\, g}f_1^{\, g}]$ evaluated using the non-evolved Gaussian TMD model of~\cite{Lansberg:2017dzg} with the evolved TMD computed along the lines described in the previous section for $M_{\Q\Q}$ = 8 GeV using the range of $b_{T\lim}$ between 2 and 8 GeV$^{-1}$. 
The main difference one can observe is the broadening of the $\qT$-spectrum when including evolution effects.
The curves are given as functions of $\qT$ in the range from 0 up to $M_{\Q\Q}$/2, to be in the validity range of TMD factorization. 

The momentum fractions of the initial gluons, $x_1$ and $x_2$, are both fixed to 10$^{-3}$.
Varying the momentum fractions does not have any significant impact on the shape of the $\qT$-spectrum or the azimuthal asymmetries. The size of the asymmetries varies by a few percent with $x$. As such variations do not change the conclusions of our analysis, we will keep the values $x_1=x_2=10^{-3}$ throughout this paper. 
This is also convenient for an experimental study, as a binning of the data in $Y_{\Q\Q}$ is not necessary to be able to compare them with predictions. 

In Fig.\ \ref{fig:qtCf1f1} (b), we show evolved results
for  $M_{\Q\Q}$ = 12, 20 and  30 GeV within the same $b_{T\lim}$ range as Fig.\ \ref{fig:qtCf1f1} (b). The broadening of the $\qT$-spectrum for increasing $M_{\Q\Q}$ is then explicit.

The azimuthal asymmetries presented in Eq.\ \eqref{asym_exp} depend on a rather complex ratio of TMD convolutions and hard-scattering coefficients. 
In the case of $J/\psi$-pair production, these expressions simplify for several reasons. 
The first one, already mentioned previously, is that because $F_2$ is small, the denominator can be approximated to be $F_1 \mathcal{C}\Big[f_1^{\, g}f_1^{\, g}\Big]$. 
Moreover, because of the symmetry of the final state, one finds the coefficients $F_3$ and $F_3'$ to be equal, simplifying the numerator of $\langle \cos(2\phi_{CS}) \rangle$ to be $F_3\Big(\mathcal{C}\Big[w_3\,f_1^{\, g}h_1^{\perp\, g}\Big]+\mathcal{C}\Big[w_3'\,h_1^{\perp\, g}f_1^{\, g}\Big]\Big)$. 
Finally, when one takes the initial-parton-momentum fractions to be equal, i.e.\ $x_1=x_2$, these two convolutions become equal as well. 
Since the $\qT$-dependence of the cross-section is contained inside the convolutions, the $\qT$-dependence of the asymmetries can be studied via the convolution ratios $\mathcal{C}\Big[w_3\,f_1^{\, g}h_1^{\perp\, g}\Big]/\mathcal{C}\Big[f_1^{\, g}f_1^{\, g}\Big]$ and $\mathcal{C}\Big[w_4\,h_1^{\perp\, g}h_1^{\perp\, g}\Big]/\mathcal{C}\Big[f_1^{\, g}f_1^{\, g}\Big]$ for $\langle \cos(2\phi_{CS}) \rangle$ and $\langle \cos(4\phi_{CS}) \rangle$, respectively.

The difference between both convolutions depends on the kind of TMDs they contain, but also the type of Bessel function generated by the angular integral and the weights. 
Because $\tilde{h}_1^{\perp\, g}$ is of order $\alpha_s$, it is naturally suppressed in comparison to $f_1^{\, g}$. 
Moreover, $\alpha_s(\mu_b)$ is growing with $b_T$ (up to its bound $\alpha_s(b_0/b_{T\max})$) and $\tilde{h}_1^{\perp\, g}$ is also broader in $b_T$ than $f_1^{\, g}$. 
The presence of $\tilde{h}_1^{\perp\, g}$ in a given convolution therefore contributes to reduce the magnitude of the integrand, and to its $b_T$-broadening. 
These effects contribute to strongly suppress \Cwtwohh with respect to \Cff. 
\Cwtwohh\ is of order $\alpha_s^2$ and its integrand is significantly broadened in $b_T$, meaning it falls faster than \Cff with increasing $\qT$. 
Indeed, as a consequence of the $b_T$-broadening, more oscillations of the $J_0$ Bessel function occur in the integrand of \Cwtwohh\ than of \Cff, before being dampened by the Sudakov factors at large $b_T$. 
Each additional oscillation in the integrand brings the convolution value closer to zero. 
More oscillations are packed in a given $b_T$-range when $\qT$ increases, widening the gap between the two convolutions, and effectively making the ratio fall with $\qT$. 
This additional effect renders the $F_2 \, \mathcal{C}\Big[w_2\,h_1^{\perp\, g}h_1^{\perp\, g}\Big]$ term truly negligible in the cross-section for $J/\psi$-pair production. 
It also means that in other processes where the hard-scattering coefficient $F_2$ may be large, the convolution itself would remain relatively small at scales larger than a few GeV. Besides, its influence on the cross-section will be strongest at the smallest TM.

\begin{figure*}[hbt!]
\centering
\subfloat[]{\includegraphics[width=\columnwidth]{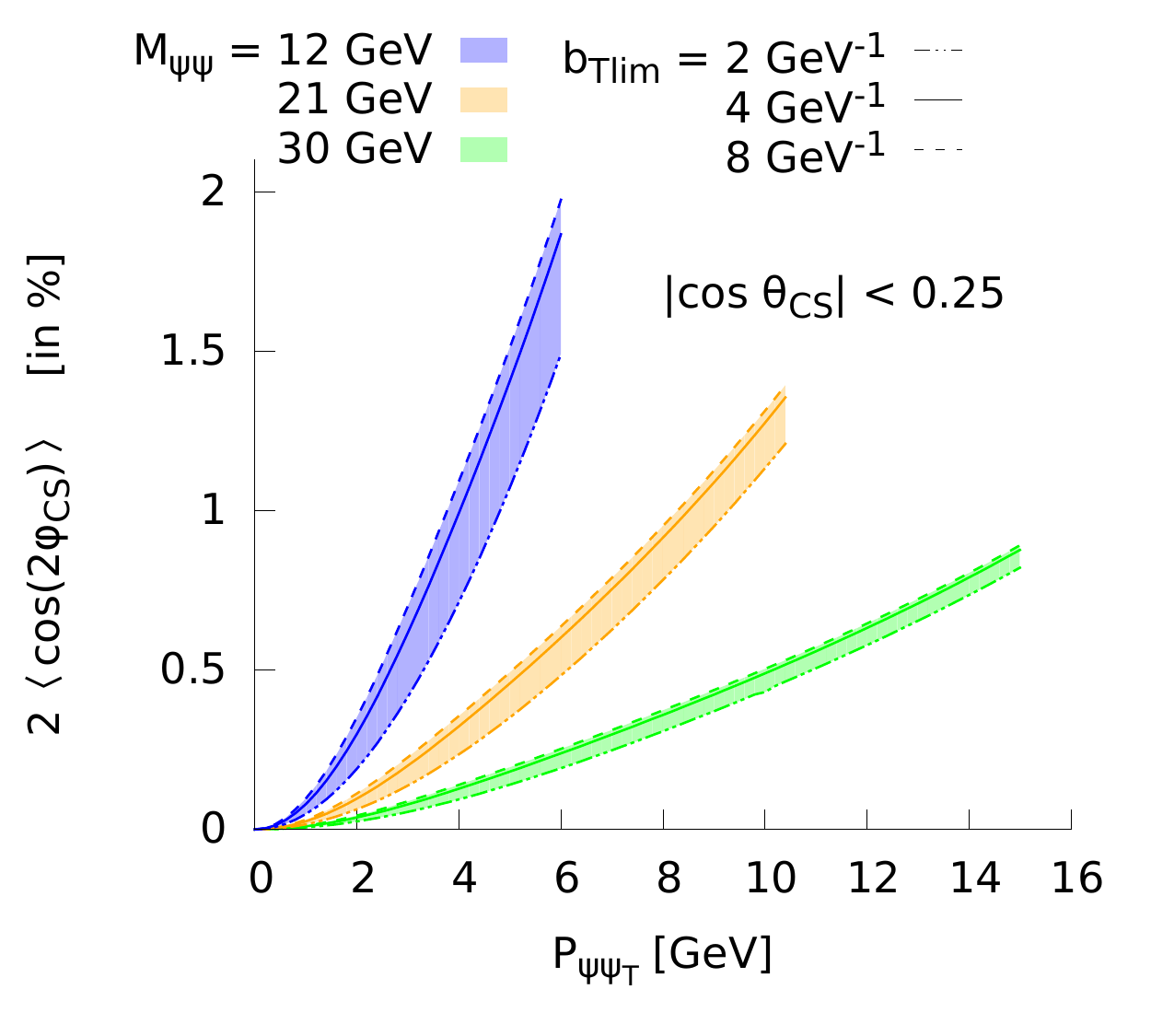}\label{fig:R2_central}}
\subfloat[]{\includegraphics[width=\columnwidth]{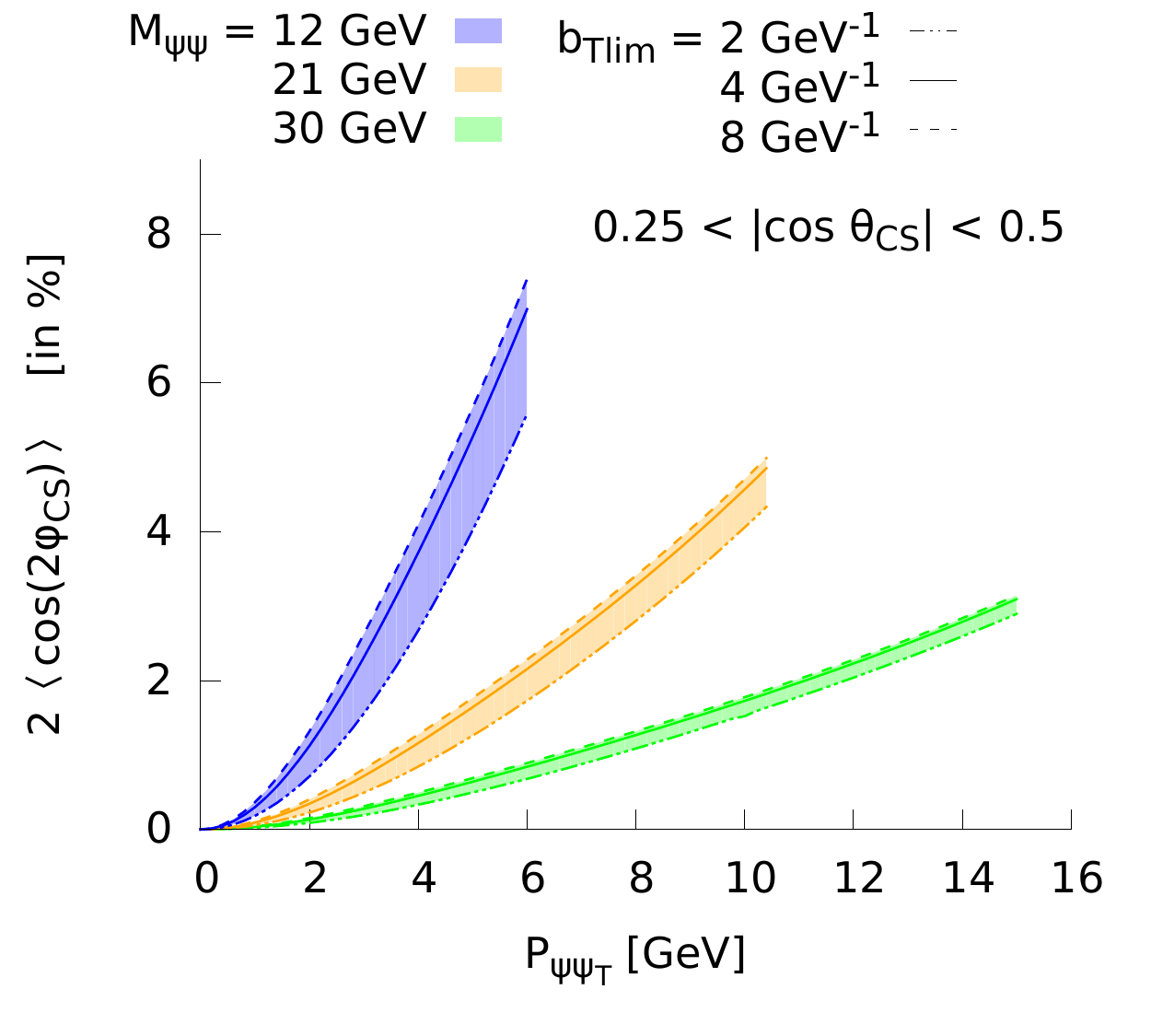}\label{fig:R2_forward}}\\
\subfloat[]{\includegraphics[width=\columnwidth]{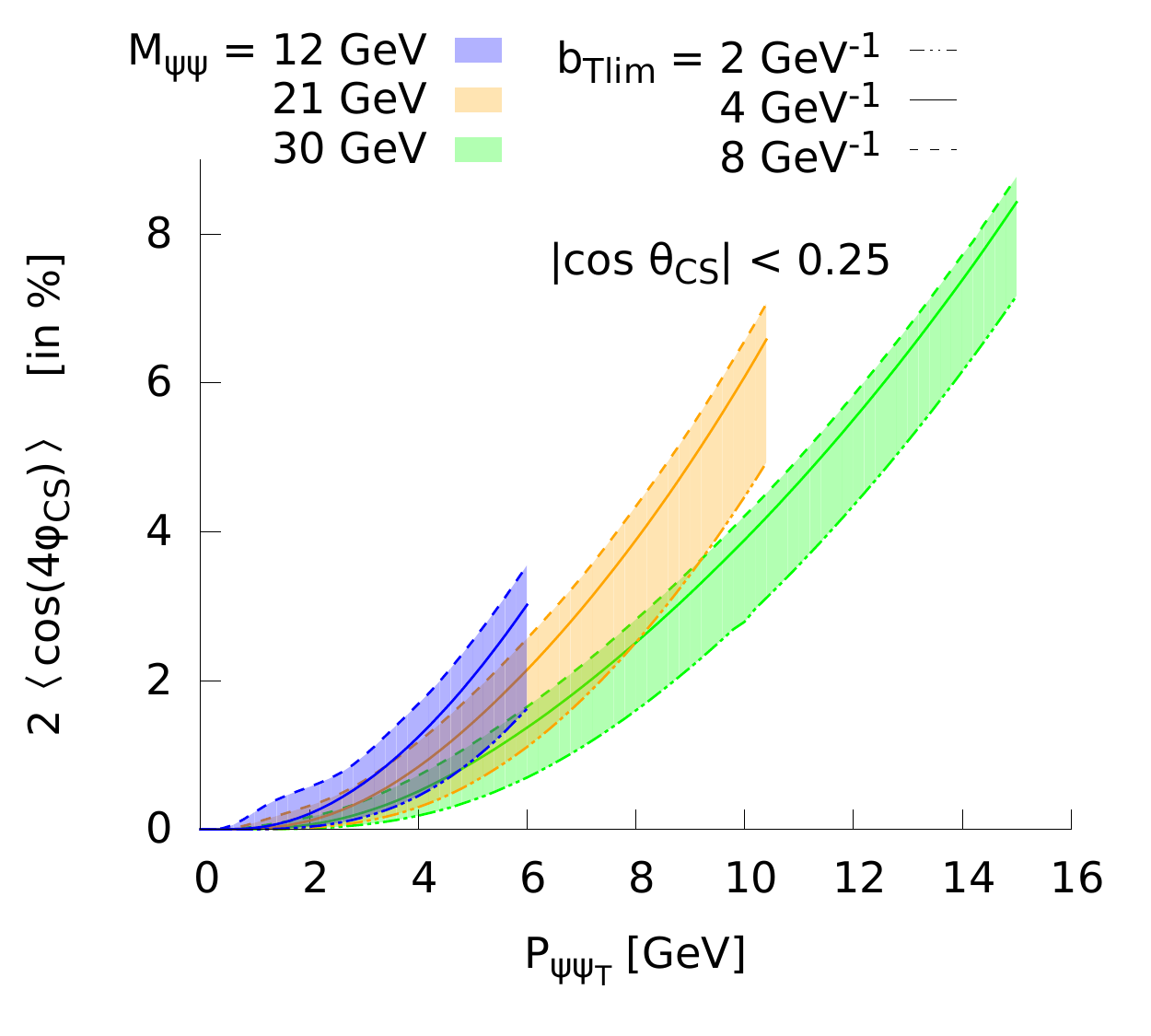}\label{fig:R4_central}}
\subfloat[]{\includegraphics[width=\columnwidth]{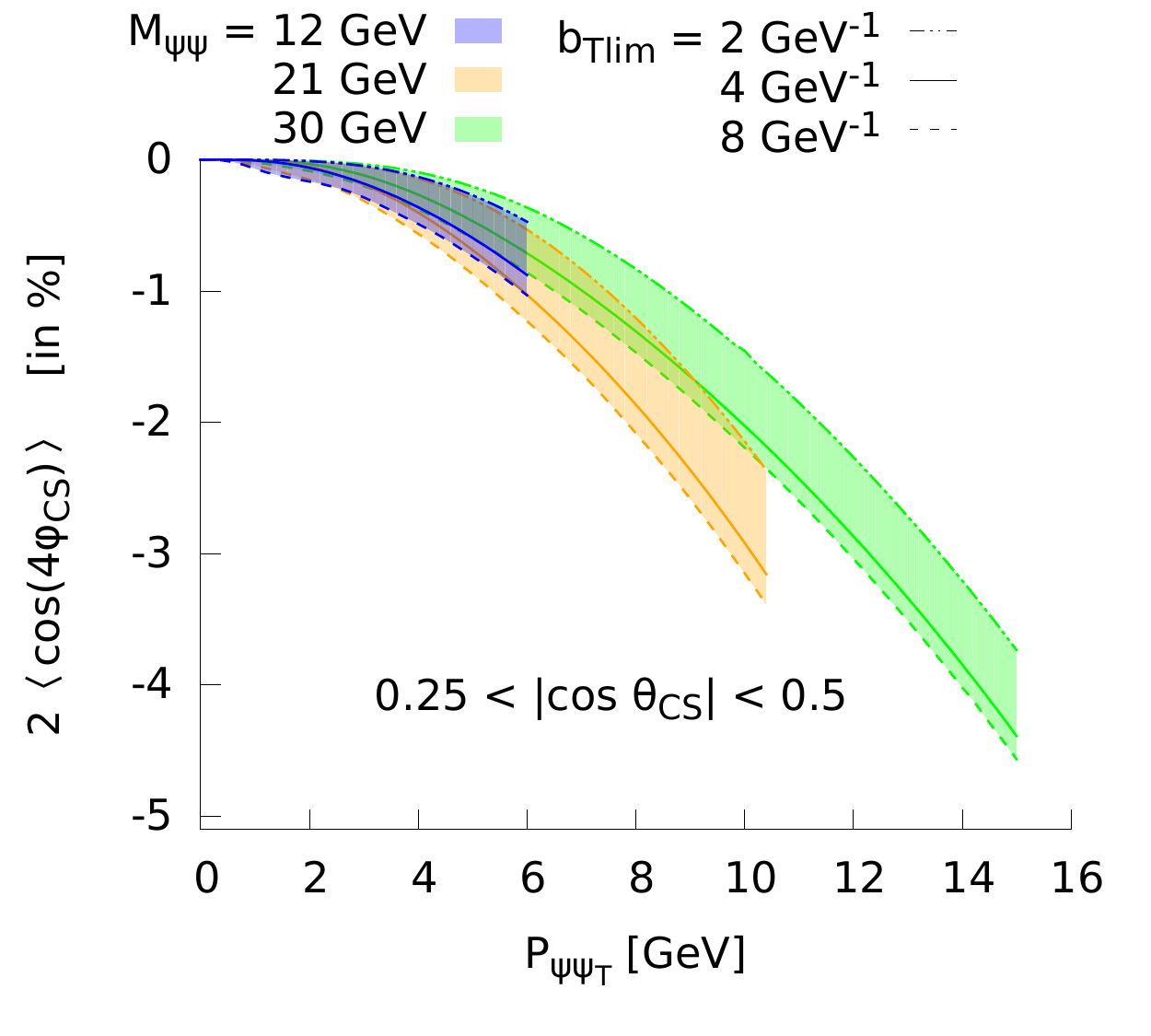}\label{fig:R4_forward}}
\caption{
The azimuthal asymmetries for di-$J/\psi$ production as functions of $\qT$. The different plots show $2\langle \cos(2\phi_{CS}) \rangle$ (a,b) and $2\langle \cos(4\phi_{CS}) \rangle$ (c,d), at $|\cos(\theta_{CS})|<0.25$ (a,c) and at $0.25<|\cos(\theta_{CS})|<0.5$ (b,d).
Results are presented for $M_{\psi\psi}$ = 12, 21 and 30 GeV, and for $b_{T\lim}$ = 2, 4 and 8 GeV$^{-1}$.}
\label{fig:asym}
\end{figure*}

The situation is different for the azimuthal asymmetries, which involve convolutions in the numerator that contain either the $J_2$ or $J_4$ Bessel functions. 
Such functions are 0 at $b_T$=0 and then grow in magnitude. 
The consequence is that the $b_T$-integrals containing such functions benefit from unsuppressed intermediate $b_T$ values. 
At some point, undampened large-$b_T$ oscillations will bring the integral value down toward 0 in a similar way as for \Cff\ and \Cwtwohh. 
Therefore, the \Cwthreefh\ and \Cwfourhh\ convolutions first grow with $\qT$ up to a peak maximum, and then decrease in value like \Cff\ does. 
Another crucial difference is that the envelopes of $J_2$ and $J_4$ tend slower toward 0 than the $J_0$ one with increasing $b_T$. 
The consequence is that \Cwthreefh\ and \Cwfourhh\ fall slower than \Cff\ with $\qT$. 
Hence the convolution ratios, and the azimuthal asymmetries, always grow with $\qT$, as can be seen in Fig.\ \ref{fig:asym}. 
In addition, as the large $b_T$ values are less suppressed than in \Cff, the azimuthal asymmetries are also more sensitive to the variations of the nonperturbative Sudakov $S_{{\rm NP}}$. 
The effect is more pronounced for \Cwfourhh\ since it contains $\tilde{h}_1^{\perp\, g}$ twice and a broader Bessel function.

\cf{fig:R2_forward} displays the $\cos(2\phi_{CS})$ asymmetry as a function of $\qT$ in the forward single $J/\psi$ rapidity region (larger $\cos(\theta_{CS})$) while \ref{fig:R4_central} displays the $\cos(4\phi_{CS})$ asymmetry in the central rapidity region (small $\cos(\theta_{CS})$ with $x_1\simeq x_2$). 
Such choices maximize the size of the asymmetries as the associated hard-scattering coefficients are larger in these regions, without modifying the shapes of the asymmetries in $\qT$ (see \cite{Lansberg:2017dzg} for a comparison between the two rapidity regions for each asymmetry). 
The uncertainty band associated with the width of $S_{{\rm NP}}$ narrows with increasing $M_{\Q\Q}$ as in Fig.\ \ref{fig:qtCf1f1}; the uncertainty remains larger for  $\langle \cos(4\phi_{CS}) \rangle$ as \Cwfourhh\ is more affected by $S_{{\rm NP}}$. 
The curves for $b_{T\lim}$ = 8 GeV$^{-1}$ (large dashes) are quite close to the ones using $b_{T\lim}$ = 4 GeV$^{-1}$ (solid line). 
Indeed, when $S_{{\rm NP}}$ is already significantly wider than $S_A$, an additional increase in its width will not affect the asymmetries anymore. 
Both convolutions in the ratios are larger with a wide nonperturbative Sudakov factor, yet this benefits the numerator $\Big(\mathcal{C}\Big[w_3\,f_1^{\, g}h_1^{\perp\, g}\Big]$ or $\mathcal{C}\Big[w_4\,h_1^{\perp\, g}h_1^{\perp\, g}\Big]\Big)$ more than the denominator $\Big(\mathcal{C}\Big[f_1^{\, g}f_1^{\, g}\Big]\Big)$, and the asymmetries are of a greater size for a wider $S_{{\rm NP}}$.

\begin{figure*}[hbt!]
\centering
\subfloat[]{\includegraphics[width=8cm]{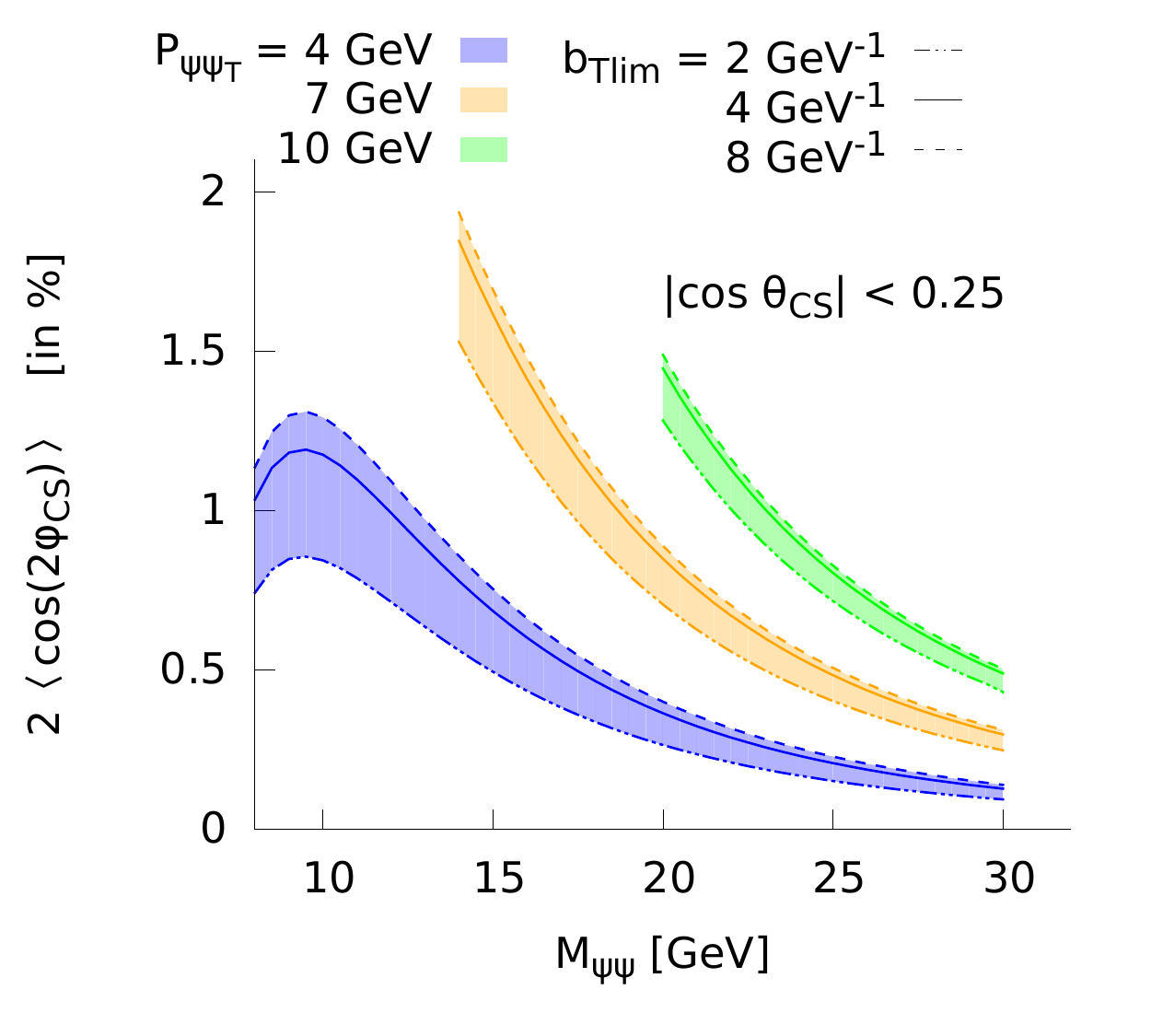}}
\subfloat[]{\includegraphics[width=8cm]{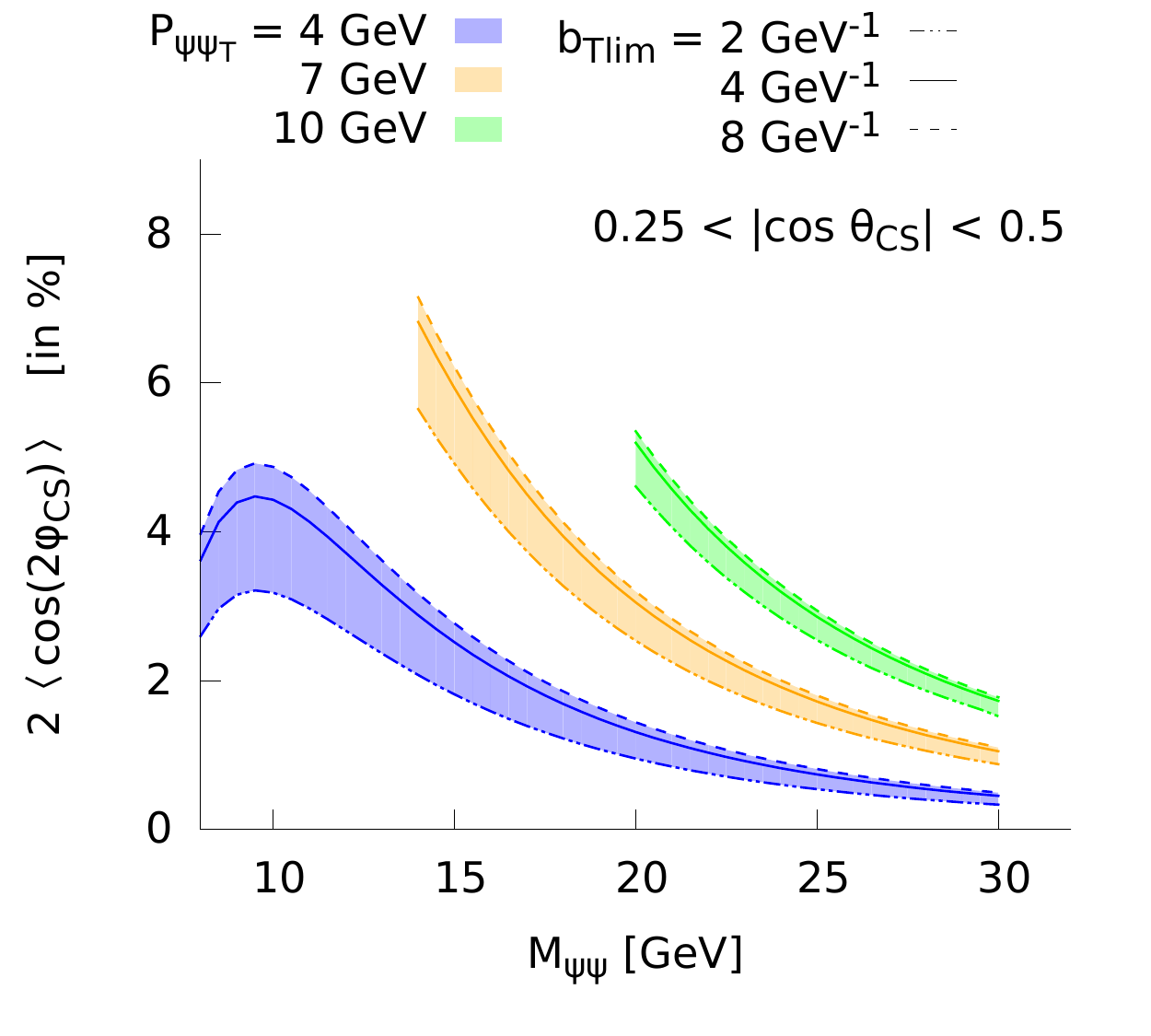}}\\
\subfloat[]{\includegraphics[width=8cm]{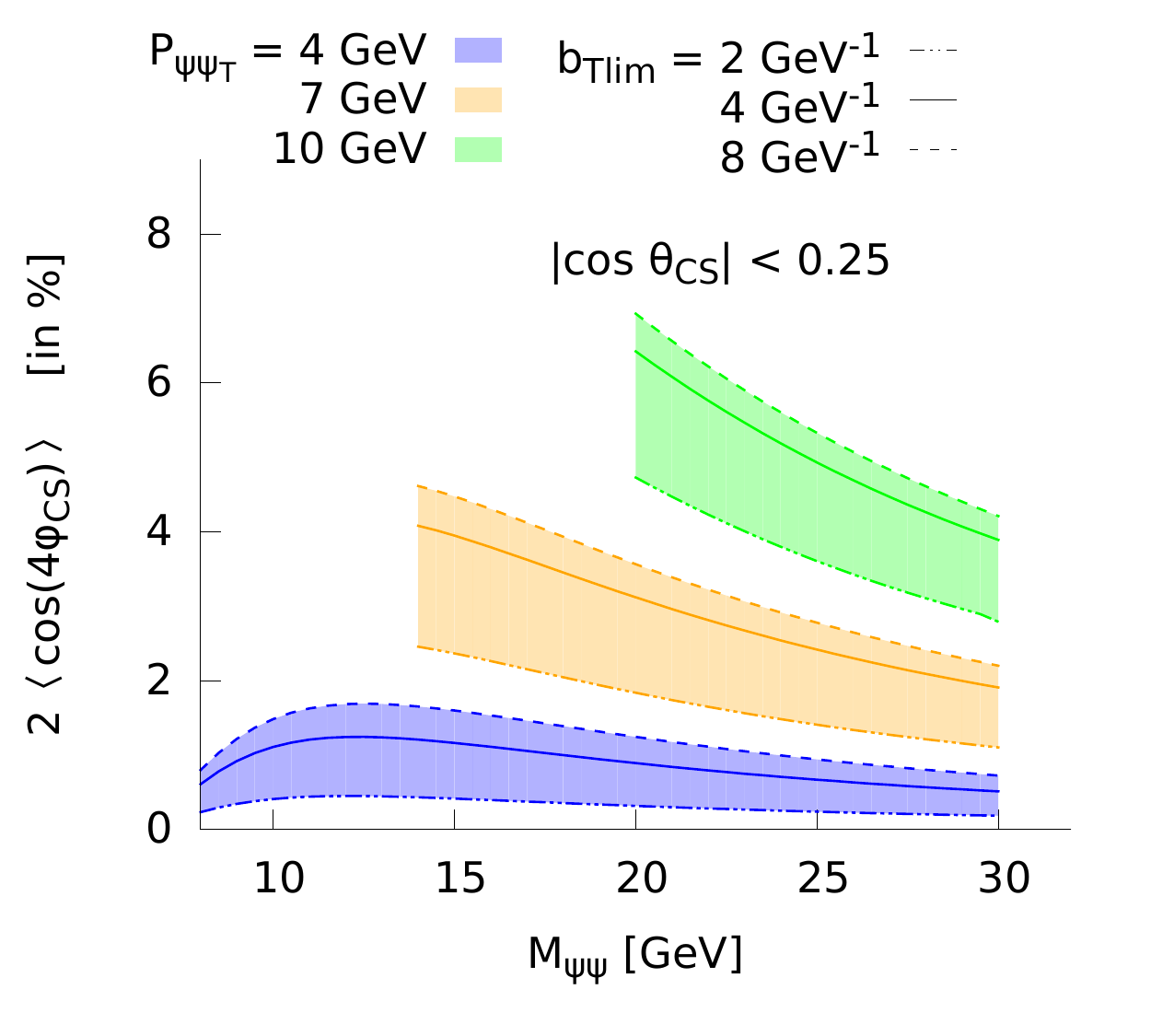}}
\subfloat[]{\includegraphics[width=8cm]{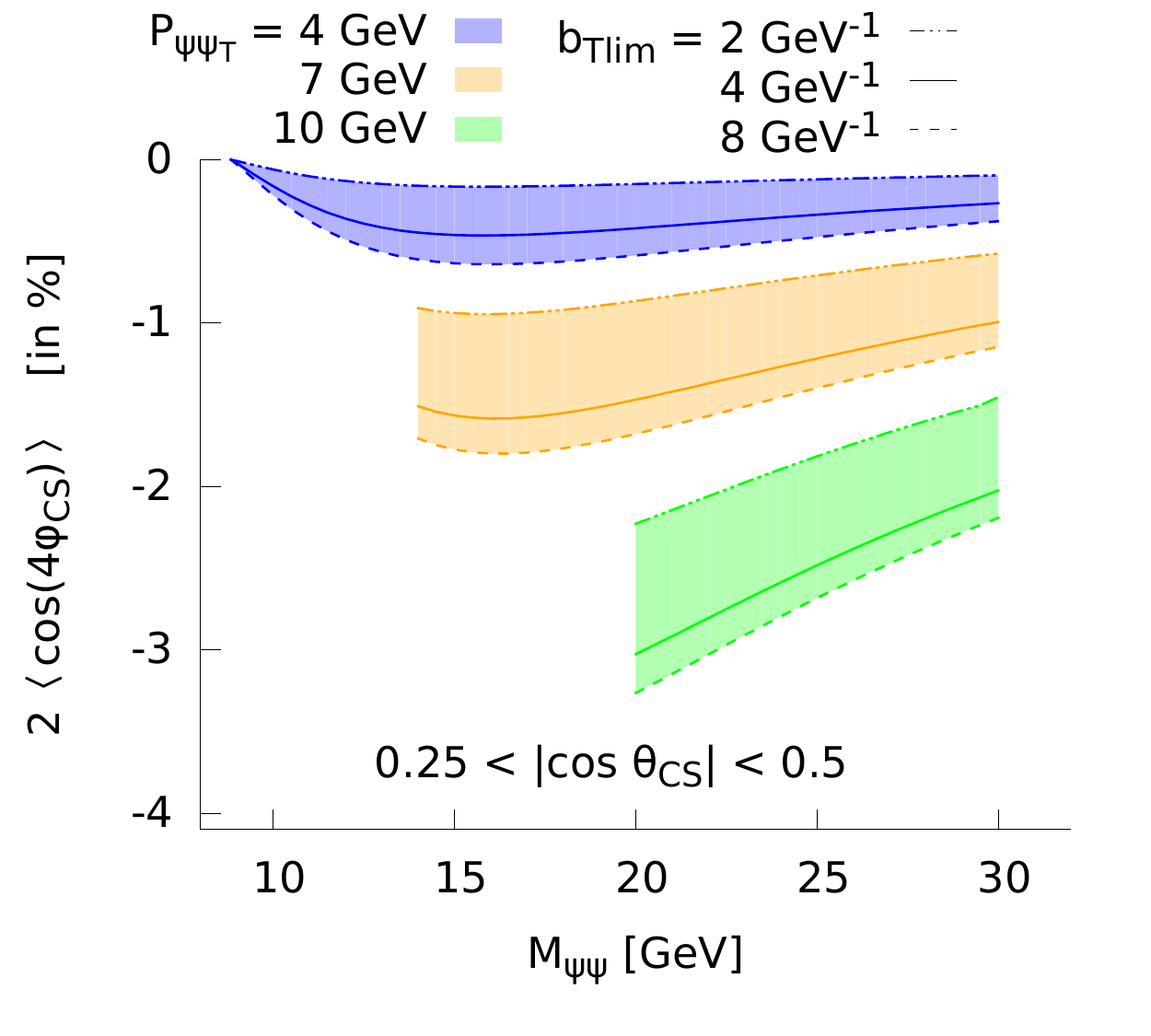}}
\caption{
The azimuthal asymmetries for di-$J/\psi$ production as functions of $M_{\psi\psi}$. The different plots show $2\langle \cos(2\phi_{CS}) \rangle$ (a,b) and $2\langle \cos(4\phi_{CS}) \rangle$ (c,d), at $|\cos(\theta_{CS})|<0.25$ (a,c) and at $0.25<|\cos(\theta_{CS})|<0.5$ (b,d). 
Results are presented for $\qT$ = 4, 7 and 10 GeV, and for $b_{T\lim}$ = 2, 4 and 8 GeV$^{-1}$.}
\label{fig:asym_Qdep}
\end{figure*}

We recall that the size of the asymmetries is also influenced by the ratio of the hard-scattering coefficients which are $M_{\Q\Q}$-dependent. 
$F_3/F_1$ peaks around $M_{\Q\Q}$ = 12 GeV which explains why the $\cos(2\phi_{CS})$ asymmetry is largest near this value. 
As discussed in \cite{Lansberg:2017dzg}, the ratio $F_4/F_1$ keeps growing with $M_{\Q\Q}$, approaching 1 at sufficiently large values. 
Yet the $\cos(4\phi_{CS})$ asymmetry gets smaller with larger $M_{\Q\Q}$. 
This can be better seen in Fig.\ \ref{fig:asym_Qdep} which depicts the same asymmetries as functions of $M_{\Q\Q}=M_{\psi\psi}$.

One first observes that, at large $M_{\Q\Q}$, the growth of the asymmetries with $\qT$ is slower. 
Indeed, in such a situation, the Sudakov factors broaden the $\qT$-shapes of the convolutions, hence the ratio varies slower. 
This slower increase is compensated by the fact that larger values of $M_{\Q\Q}$ allow for an extended growth of the asymmetry over a greater $\qT$-range of validity for the TMD formalism. 
Secondly, the convolution ratios at a fixed value of $\qT$ also evolve with $M_{\Q\Q}$. 
The computable $M_{\Q\Q}$-dependence is encoded in the perturbative Sudakov factor $S_A$, while $S_{{\rm NP}}$ is also logarithmically varying with $M_{\Q\Q}$ \cite{Collins:1981va}. 
Both $S_A$ and $S_{{\rm NP}}$ get narrower in $b_T$ with increasing $M_{\Q\Q}$, leading to a decrease of the value of the convolutions. 
\Cwthreefh\ and \Cwfourhh\ are more sensitive to the large $b_T$-value dampening and therefore fall faster with $M_{\Q\Q}$ than \Cff. 
This results in decreasing convolution ratios, with a steeper fall for \Cwfourhh. 
However the azimuthal asymmetries also depend on the evolution with $M_{\Q\Q}$ of the hard-scattering coefficients ratios. 
Since $F_4/F_1$ keeps growing while $F_3/F_1$ falls after peaking at $M_{\Q\Q} \simeq$ 12 GeV, $\langle \cos(4\phi_{CS}) \rangle$ will actually decrease slower than $\langle \cos(2\phi_{CS}) \rangle$.

The large variations of the width of $S_{{\rm NP}}$ generate moderate uncertainties on the size of the asymmetries. 
The latter, although consequently smaller than when computed in a bound-saturating model \cite{Lansberg:2017dzg}, still reach reasonable sizes, up to 5\%-10\%. 
We used the same nonperturbative Sudakov factor for all TMD convolutions in these computations, but the $M_{\Q\Q}$-independent part is actually expected to be non-universal. 
We checked that individually changing the width of $S_{{\rm NP}}$ within the $b_{T\lim}$-ranges used in this study inside the different types of convolutions, does not bring any significant modification on the observables. 

So far, there are still no experimental data allowing for an extraction of the gluon TMDs inside unpolarized protons. 
We believe that the numerous $J/\psi$-pair-production events recorded at the LHC can give us access to information about the nonperturbative components of $f_1^{\, g}$ and $h_1^{\perp\, g}$, provided the events are selected with kinematics within the validity range of TMD factorization, $\qT<M_{\Q\Q}/2$.

\subsection{$\Upsilon$-pair production}

It is also of interest to look at $\Upsilon$-pair production. 
The partonic subprocess is identical to that of di-$J/\psi$ production. 
In the non-relativistic limit, where $M_\Upsilon=2 m_b$, the main difference comes from 
the mass of the heavy quark. We note that the value of the 
non-relativistic wave function at the origin (or equivalently the NRQCD LDME for the CS transition) also differs but cancels in the ratios which we consider. The feed-down pattern is also clearly different. However, as announced, we will neglect the resulting (small) feed-down effects.

Owing to this larger mass, such a process probes the evolution at generally higher scales. 
The coupling constant $\alpha_s$ is also smaller which increases the precision of the perturbative expansion. 
Higher scales also mean that the process is less sensitive to the (large $b_T$) nonperturbative behavior  of the gluon TMDs. Hence, it is also less affected by the uncertainties associated with this unconstrained component. 

\begin{figure*}[h]
\centering 
\subfloat[]{\includegraphics[width=\columnwidth]{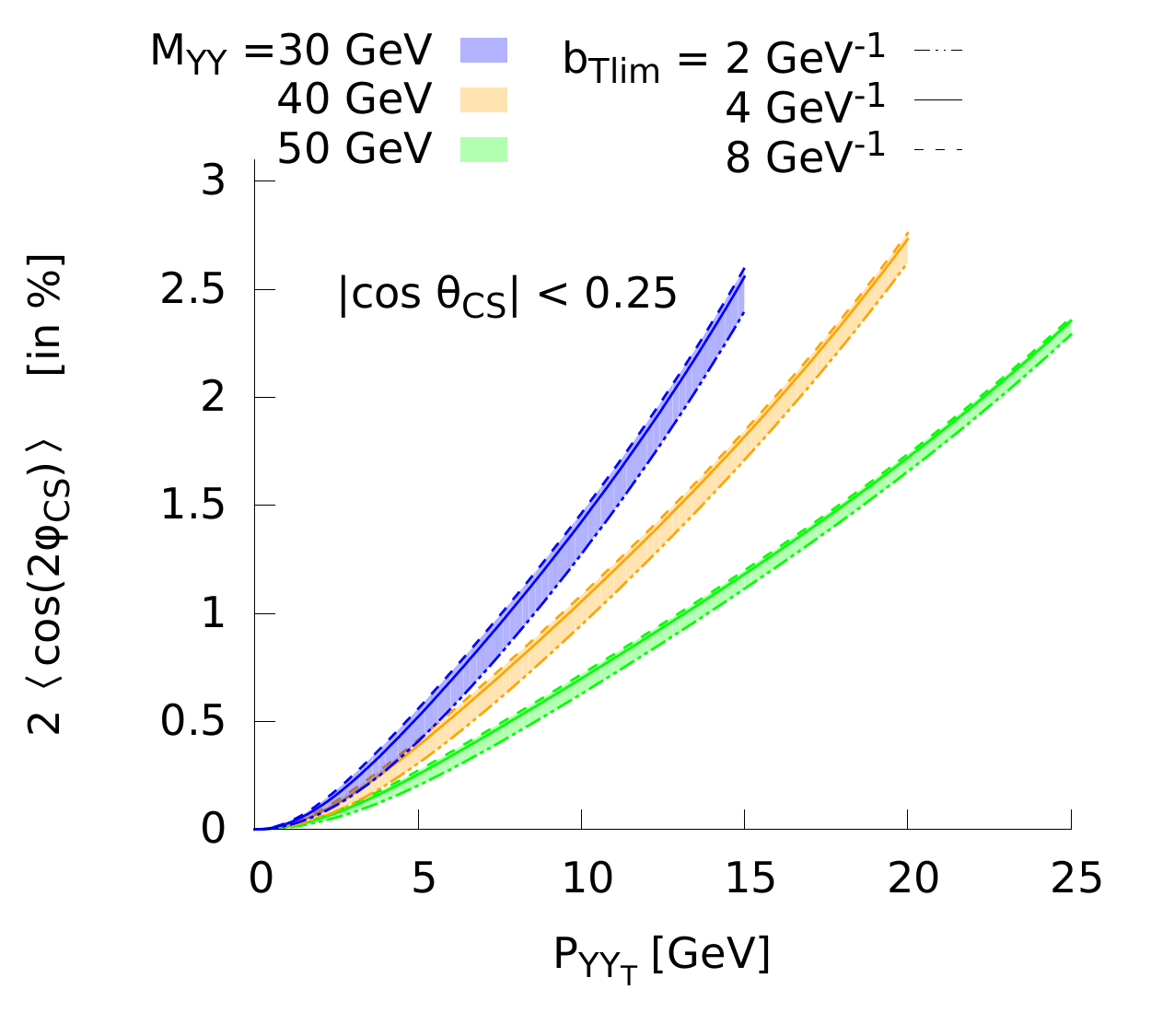}\label{fig:R2_central_Ups}}
\subfloat[]{\includegraphics[width=\columnwidth]{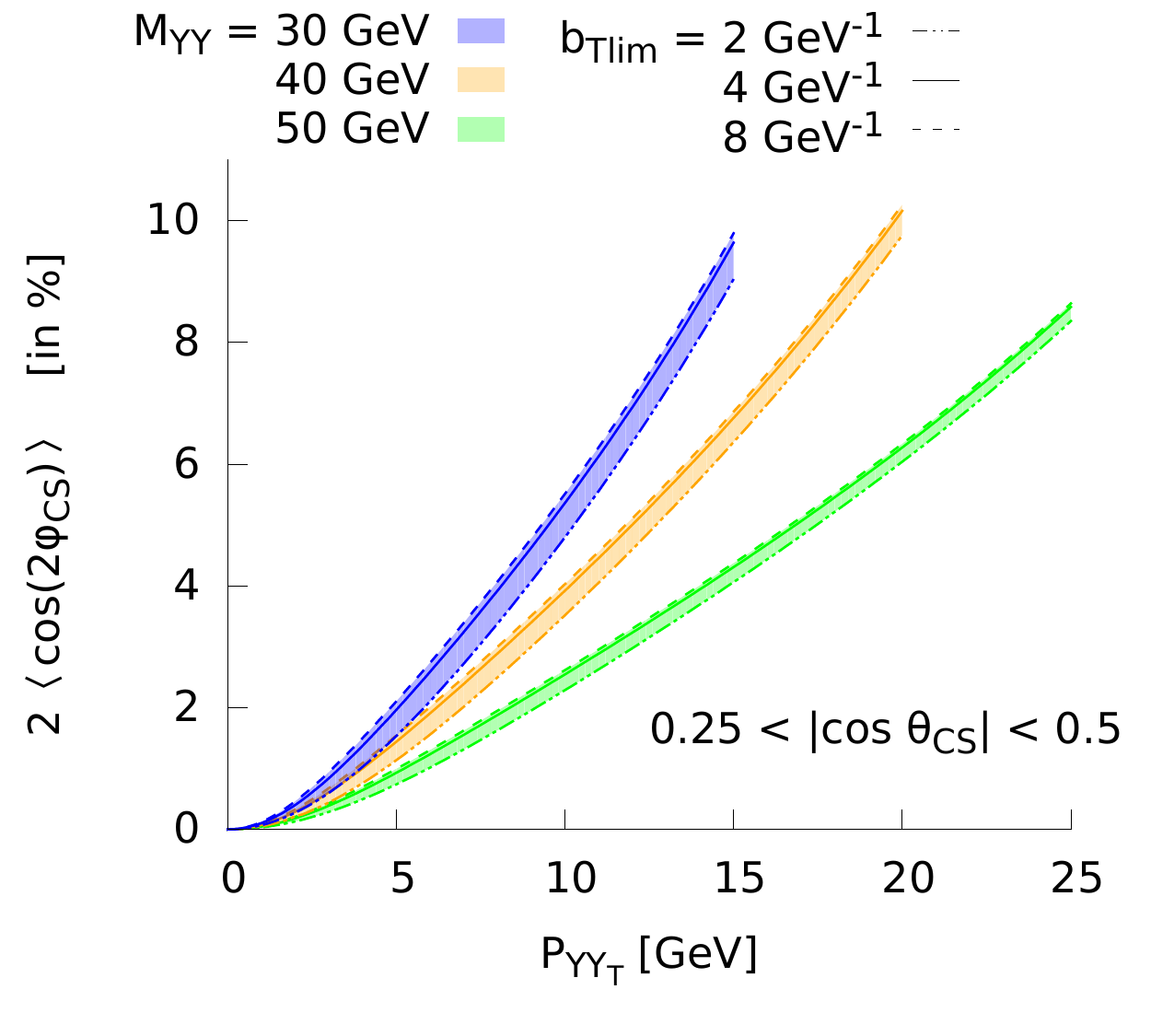}\label{fig:R2_forward_Ups}}\\
\subfloat[]{\includegraphics[width=\columnwidth]{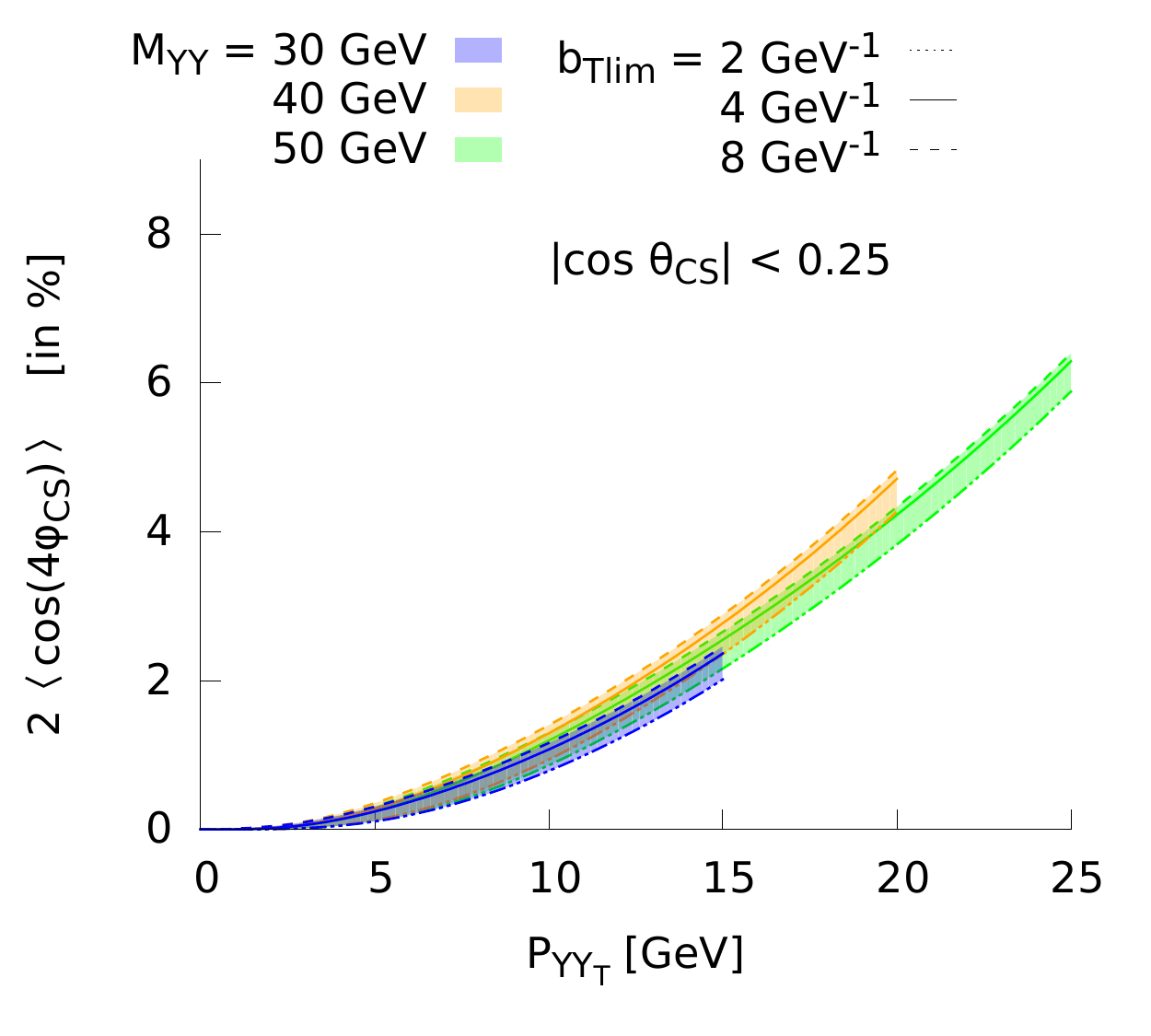}\label{fig:R4_central_Ups}}
\subfloat[]{\includegraphics[width=\columnwidth]{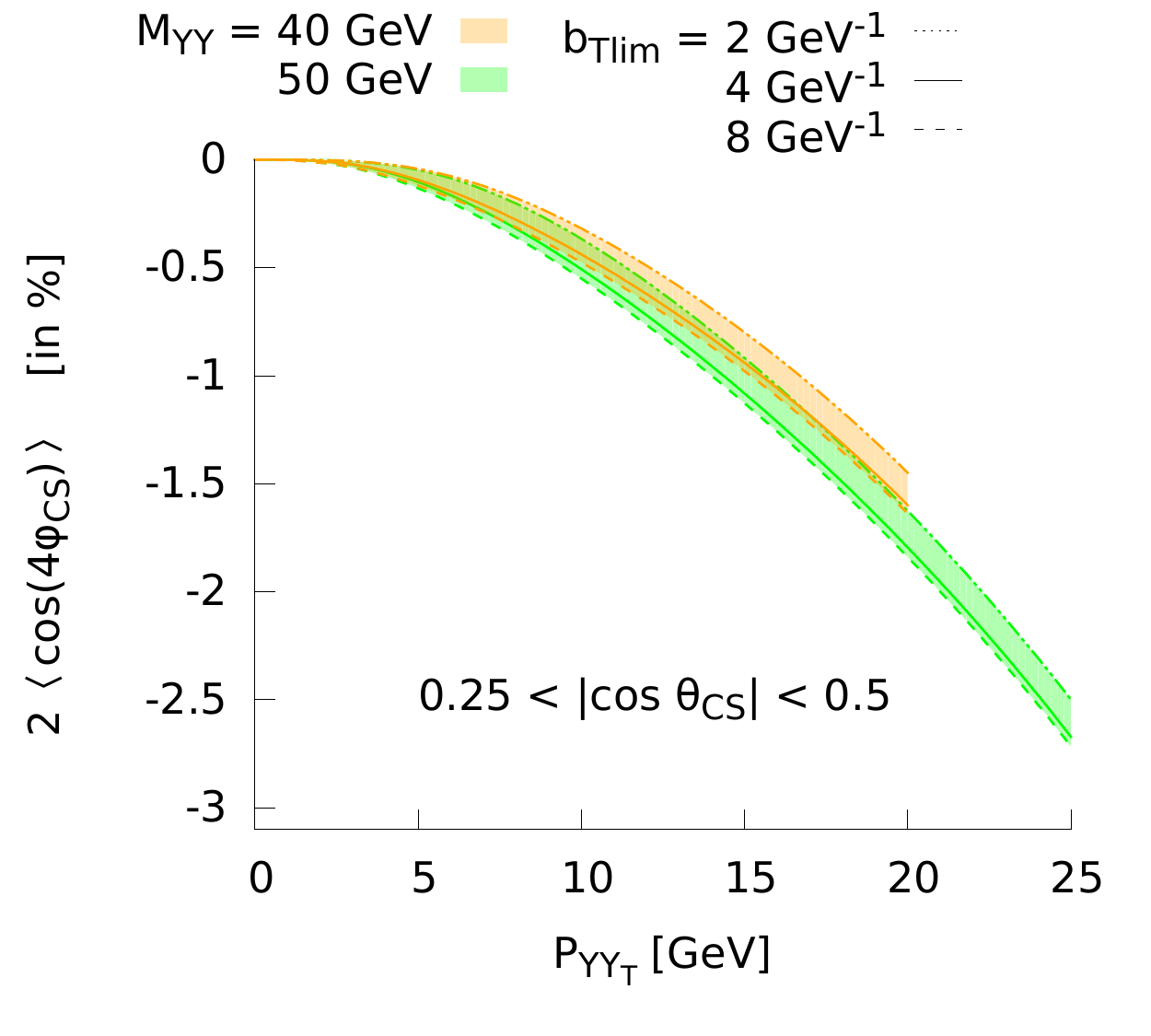}\label{fig:R4_forward_Ups}}
\caption{
The azimuthal asymmetries for di-$\Upsilon$ production as functions of $\qT$. The different plots show $2\langle \cos(2\phi_{CS}) \rangle$ (top) and $2\langle \cos(4\phi_{CS}) \rangle$ (bottom), at $|\cos(\theta_{CS})|<0.25$ (left) and at $0.25<|\cos(\theta_{CS})|<0.5$ (right).
Results are presented for $M_{\Upsilon\Upsilon}$ = 30, 40 and 50 GeV, and for $b_{T\lim}$ = 2, 4 and 8 GeV$^{-1}$. 
Results for $M_{\Upsilon\Upsilon}$ = 30 GeV are not included in (d) as they are below percent level.}
\label{fig:asym_Ups}
\end{figure*}

On the experimental side, $\Upsilon$-pair production is admittedly a rare process. Yet, it starts to be accessible at the LHC. The first analysis by the CMS collaboration at $\sqrt{s}$ = 8 TeV only comprised a 40-event sample \cite{Khachatryan:2016ydm} but a second one is forthcoming. 
During the future high luminosity LHC runs, it will definitely be possible to record a sufficient number of events for a TMD analysis of both the $\qT$ and azimuthal dependences of the yield. 
Fig.\ \ref{fig:asym_Ups} depicts the azimuthal modulations for $\Upsilon$-pair production as functions of $\qT$ up to $M_{\Q\Q}/2$, for values of $M_{\Q\Q}$ of 30, 40 and 50 GeV.

The uncertainty bands associated with the width of $S_{{\rm NP}}$ are clearly narrower than in the $J/\psi$ case. 
The $\cos(2\phi_{CS})$ asymmetry in Fig.\ \ref{fig:R2_forward_Ups} reaches 10\% at $M_{\Q\Q}$ = 40 GeV, which is the value for which the corresponding hard-scattering coefficient ratio $F_3/F_1$ peaks for $\Upsilon$-pair production. 
Moreover, the decrease of the hard-scattering coefficient past the peak is slower, allowing the asymmetry to remain of similar size at $M_{\Q\Q}$ = 40 and 50 GeV. 

\section{Conclusions}

In this paper we discussed the potential of double $J/\psi$ and $\Upsilon$ production for the study of the gluon TMDs inside unpolarized protons at the LHC. 
We presented the advantages of quarkonia as probes of these TMDs.
We improved on previous results \cite{Lansberg:2017dzg} by including TMD evolution effects, rendering the results more realistic and effectively taking into account QCD corrections that describe the evolution with the invariant mass $M_{\Q\Q}$ of the quarkonium pair.
We used a simple $b_T$-Gaussian of variable width to parametrize the nonperturbative Sudakov factor $S_{{\rm NP}}$ in order to estimate how important its impact is on the predicted yield and asymmetries, as it currently remains unconstrained in the gluon case.

We discussed the broadening of the $\qT$-spectrum due to the evolution in the case of double-$J/\psi$ production, as well as the uncertainty associated with a variation of the width of $S_{{\rm NP}}$ between 2 and 8 GeV$^{-1}$. {As expected, 
we found} that its influence decreases at large $M_{\Q\Q}$ as the perturbative component of the TMDs becomes dominant. 
We also computed the $\langle \cos(2,\!4\phi_{CS}) \rangle$ asymmetries as functions of $\qT$ and $M_{\Q\Q}$. 
We found a notable suppression of the asymmetries in comparison to \cite{Lansberg:2017dzg}, caused by the fact that $h_1^{\perp\, g}$ appears at order $\alpha_s$ in the evolution formalism. 
We nevertheless found that such asymmetries still reach reasonable sizes for larger $\qT$ values and could be observed in the events already collected and to be recorded in the future. 
We found that the size of the asymmetries increases with $\qT$. 
Such a behavior is explained by the relative slower fall in $\qT$ of the TMD convolutions containing $h_1^{\perp\, g}$. 

TMD factorization needs to be matched onto its collinear counterpart when $\qT$ approaches $M_{\Q\Q}$. 
Since the latter generates no asymmetries at leading twist, a $Y$-term becomes necessary at some point in order to neutralize the growth of the asymmetries and force them toward zero. 
We also observed that, in spite of the hard-scattering coefficient ratio $F_4/F_1$ approaching 1 at large energy, the $\cos(4\phi)$ asymmetry actually falls with $M_{\Q\Q}$. 

Overall we conclude that $J/\psi$-pair production is a promising process to measure azimuthal asymmetries related to gluon TMDs as well as the effect of the evolution on the $\qT$-spectrum. 
The energy threshold for this process is relatively low, making it sensitive to the nonperturbative component of the TMDs. 
The large event sample to be collected by the different collaborations at the LHC should give enough statistics to constrain them. 
$\Upsilon$-pair production presents the interesting opportunity to measure sizeable asymmetries at scales where perturbative contributions dominate, with a reduced necessity to include higher-order corrections. 
We also presented predictions for the asymmetries as functions of $\qT$ for $\Upsilon$-pair production. 
With sufficient data to come, it would allow for a complementary extraction of the gluon TMDs, while the expected size of asymmetries remain similar. 
Although $\Upsilon$ pairs remain extremely rare at the LHC, the future high-luminosity runs will make it possible to acquire enough statistics.

Accessing information about the gluon TMDs can thus already be done at the LHC using quarkonium production, although more efforts in the direction of \cite{Echevarria:2019ynx} are needed in order to obtain rigorous factorization theorems and expressions beyond tree level.
It would give us a preview of what we can expect to find at a future Electron-Ion Collider \cite{Accardi:2012qut} or fixed-target experiments at the LHC \cite{Hadjidakis:2018ifr,Lansberg:2015lva,Massacrier:2015qba,Lansberg:2012kf,Brodsky:2012vg}, where these distributions should be accessible through different reactions. 
Because of the fundamental differences in these experimental setups, it is of great interest to measure the same TMDs using both of them, in order to be able to check fundamental predictions of the formalism such as the evolution and the universality.

\section*{Acknowledgements}
{The work of MS was in part supported within the framework of the TMD Topical Collaboration and
that of FS and JPL by the CNRS-IN2P3 project TMD@NLO. This project is also supported by the European Union's Horizon 2020 research and innovation programme under grant agreement No 824093. MGE is supported by the Marie Sk\l odowska-Curie grant \emph{GlueCore} (grant agreement No. 793896).}

\bibliographystyle{utphys}

\providecommand{\href}[2]{#2}\begingroup\raggedright\endgroup

\clearpage 

\end{document}